\documentclass[apj,numberedappendix,appendixfloats]{emulateapj}
\usepackage{apjfonts}
\usepackage{natbib}
\usepackage{amssymb}
\usepackage{enumitem}
\usepackage{braket}
\usepackage[breaklinks,colorlinks,citecolor=blue,linkcolor=magenta]{hyperref}

\newcommand{\etal}{et~al.}
\newcommand{\MgIIdblt}{{\rm Mg}\kern 0.1em{\sc ii}~$\lambda\lambda 2796, 2803$}
\newcommand{\MgII}{\hbox{{\rm Mg}\kern 0.1em{\sc ii}}}
\newcommand{\NII}{\hbox{{\rm N}\kern 0.1em{\sc ii}}}
\newcommand{\HI}{\hbox{{\rm H}\kern 0.1em{\sc i}}}
\newcommand{\CIV}{\hbox{{\rm C}\kern 0.1em{\sc iv}}}
\newcommand{\OII}{\hbox{{\rm O}\kern 0.1em{\sc ii}}}
\newcommand{\OIII}{\hbox{{\rm O}\kern 0.1em{\sc iii}}}
\newcommand{\OVI}{\hbox{{\rm O}\kern 0.1em{\sc vi}}}
\newcommand{\OV}{\hbox{{\rm O}\kern 0.1em{\sc v}}}
\newcommand{\OVII}{\hbox{{\rm O}\kern 0.1em{\sc vii}}}
\newcommand{\Lya}{\hbox{{\rm Ly}\kern 0.1em $\alpha$}}
\newcommand{\logtenmv}{\hbox{$\log(M_{\rm h}/M_\odot)$}}
\newcommand{\kms}{\hbox{km~s$^{-1}$}}

\shorttitle{\sc ~Halo Mass Dependence of {\OVI} Kinematics}
\shortauthors{\sc Ng {\etal}}

\begin{document}

\title{Kinematics of the {\OVI} Circumgalactic Medium: Halo Mass
  Dependence and \\Outflow Signatures}

\author{
Mason Ng$^{1,2,}$\altaffilmark{$\dagger$},
Nikole M.~Nielsen$^{1,3}$,
Glenn G.~Kacprzak$^{1,3}$,
Stephanie K.~Pointon$^{1,3}$,
Sowgat Muzahid$^{4,5}$,\\
Christopher W.~Churchill$^{6}$,
and
Jane C.~Charlton$^{4}$,
}

%\altaffiltext{1}{Department of Physics and Kavli Institute for
%Astrophysics and Space Research, Massachusetts Institute of
%Technology, Cambridge, MA 02139, USA; email: masonng@mit.edu}

\affil{$^1$ Centre for Astrophysics and Supercomputing, Swinburne
University of Technology, Hawthorn, Victoria 3122, Australia\\
$^2$ Research School of Astronomy and Astrophysics, Australian
National University, ACT 2611, Australia\\
$^3$ ARC Centre of Excellence for All Sky Astrophysics in 3 Dimensions
(ASTRO 3D), Australia \\
$^4$ Department of Astronomy \& Astrophysics, The Pennsylvania State
University, State College, PA 16801, USA\\
$^5$ Leiden Observatory, Leiden University, PO Box 9513, NL-2300 RA
Leiden, The Netherlands\\
$^6$ Department of Astronomy, New Mexico State University, Las
Cruces, NM 88003, USA
}

\altaffiltext{$\dagger$}{masonng@mit.edu}

\begin{abstract}

We probe the high-ionization circumgalactic medium by examining
absorber kinematics, absorber--galaxy kinematics, and average
absorption profiles of 31 {\OVI} absorbers from the ``Multiphase
Galaxy Halos'' Survey as a function of halo mass, redshift,
inclination, and azimuthal angle. The galaxies are isolated at
$0.12<z_{\rm gal}<0.66$ and are probed by a background quasar within
$D\approx 200$~kpc. Each absorber--galaxy pair has {\it Hubble Space
  Telescope} images and COS quasar spectra, and most galaxy redshifts
have been accurately measured from Keck/ESI spectra. Using the
pixel-velocity two-point correlation function (TPCF) method, we find
that {\OVI} absorber kinematics have a strong halo mass
dependence. Absorbers hosted by $\sim L^{\ast}$ galaxies have the
largest velocity dispersions, which we interpret to be that the halo
virial temperature closely matches the temperature at which the
collisionally ionized {\OVI} fraction peaks. Lower mass galaxies and
group environments have smaller velocity dispersions. Total column
densities follow the same behavior, consistent with theoretical
findings. After normalizing out the observed mass dependence, we
studied absorber--galaxy kinematics with a modified TPCF and found
non-virialized motions due to outflowing gas. Edge-on minor axis gas
has large optical depths concentrated near the galaxy systemic
velocity as expected for bipolar outflows, while face-on minor axis
gas has a smoothly decreasing optical depth distribution out to large
normalized absorber--galaxy velocities, suggestive of decelerating
outflowing gas. Accreting gas signatures are not observed due to
``kinematic blurring'' in which multiple line-of-sight structures are
observed. These results indicate that galaxy mass dominates {\OVI}
properties over baryon cycle processes.

\end{abstract}


\section{Introduction}
\label{sec:intro}

The prodigious reserves of gas surrounding galaxies in the
circumgalactic medium (CGM) play an important role in galaxy evolution
\citep[see review by][]{tumlinson17}. This gas is primarily derived
from the intergalactic medium \citep[IGM, e.g.,][]{putman12, cooper15,
  glidden16}, from cannibalizing satellite galaxies
\citep[e.g.,][]{cole00, cox08, qu11, alonso12a, lambas12, kaviraj14,
  ownsworth14, gomez-guijarro18}, and from galactic feedback
\citep[e.g.,][]{strickland09, schaye15, vandevoort17, butler17,
  correa18}. The general accepted picture of how a typical galaxy
evolves includes the accretion of relatively metal-poor gas from the
CGM onto the galactic disk \citep[see review by][]{kacprzak17}, which
is used to fuel star formation. Gas is then driven out of the galactic
disk in outflows when massive stars explode as supernovae and produce
metal-enriched winds \citep[e.g.,][]{shen12, lehner13, ford14,
  muzahid15}. The velocities of the outflowing gas do not usually
exceed the escape velocity of the galaxy \citep[e.g.,][]{tumlinson11,
  bouche12, stocke13, mathes14, bordoloi14}, thus the gas is recycled
back onto the galaxy and could fuel further episodes of star formation
\citep[e.g.,][]{oppenheimer10, ford14, vandevoort17}. This paints the
picture of the baryon cycle within the galaxy virial radius.

The {\OVI} $\lambda\lambda1031,1037$ absorption doublet is a common
tracer of the CGM, particularly in the high-temperature regime of
$T\sim10^{5}\rm{\,K}$ \citep[e.g.,][]{prochaska11, tumlinson11,
  stocke13, savage14, churchill15, johnson15, kacprzak15,
  werk16}. \citet{oppenheimer16} employed the EAGLE simulations to
investigate the presence and role of different oxygen species in the
CGM, assuming that {\OVI} is collisionally ionized. They found that
{\OVI} is not the dominant oxygen species in the CGM, and that the
column densities for {\OVI} peak for $L_*$ galaxies, while dropping
for lower mass halos and group halos. This is thought to be due to the
{\OVI} ionization fraction strongly tracing the virial temperature of
the galaxy, where the associated virial temperature for $L_*$ galaxies
provides the optimal conditions for the presence of {\OVI}. For less
massive galaxies, the virial temperature would be too cool for strong
{\OVI} presence, whereas the virial temperature would be too high for
group environments as a large fraction of {\OVI} is ionized out
to higher ionization species. \citet{nelson18} found similar trends in
the {\OVI} column density with halo mass, but attributed them to black
hole feedback \citep[also see][]{oppenheimer18}. Several other works
have also shown this trend in both observations \citep{bielby19,
  zahedy19} and with gaseous halo models \citep{qu18}.

Using a sample of quasar absorption-line spectra from {\it HST}/COS
identified as part of the ``Multiphase Galaxy Halos'' Survey,
\citet{kacprzak15} found that {\OVI} has an azimuthal angle
preference, where {\OVI} tends to reside along the projected major
axis ($0^\circ \leq \Phi \leq 20^\circ$) and/or along the projected
minor axis ($60^\circ \leq \Phi \leq 90^\circ$). They also found a
very weak dependence of the {\OVI} absorption on the galaxy
inclination, where the covering fraction of the {\OVI} gas is roughly
constant over all inclination angles except for $i>70^\circ$, as the
high inclination minimizes the geometrical cross-section of gas
flows. Moreover, the mean equivalent widths of {\OVI} in lower
inclination ($i<45^\circ$) galaxies and higher inclination
($i>45^\circ$) galaxies are consistent with each other.

Previous kinematics studies examined the absorber velocity dispersions
of {\OVI} with pixel-velocity two-point correlation functions (TPCFs)
to characterize the absorber velocity dispersions for isolated
galaxies \citep{nielsen17}. The authors found that there was no
dependence of~{\OVI} kinematics on the inclination angle, azimuthal
angle, and/or galaxy color, which indirectly suggests a lack of
dependence on current star formation activity. They attribute this to
{\OVI} absorbers having ample time to mix and form a kinematically
uniform halo surrounding the galaxies. This is consistent with
\citet{ford14}, who found that {\OVI} in simulations likely traces gas
that originates from ancient outflows. These results are in contrast
to {\MgII} kinematics, which depends strongly on galaxy color,
redshift, inclination, and azimuthal angle \citep{magiicat5,
  magiicat4}.

\citet{pointon17} examined~{\OVI} kinematics using TPCFs for galaxy
group environments and found that the {\OVI} absorption profiles for
galaxy group environments are narrower compared to isolated
galaxies. They posit that the virial temperature of the CGM in galaxy
group environments (with more massive halos) is hot enough to ionize a
larger fraction of {\OVI} to higher order species to result in a lower
{\OVI} ionization fraction compared to isolated galaxies, consistent
with the findings of \citet{oppenheimer16}. This suggests that halo
mass needs to be considered when studying the absorber kinematics of
{\OVI}.

Focusing on {\OVI} absorber--galaxy kinematics, \citet{tumlinson11}
found that {\OVI} absorber--galaxy velocities rarely exceed the host
galaxy escape velocity, indicating that the gas is
bound. \citet{mathes14} found similar results, but noted that the
fraction of gas that exceeds host galaxy escape velocities decreases
with increasing halo mass. The authors suggested that wind recycling
is increasingly important as the halo mass increases, consistent with
simulations \citep{oppenheimer10}. Most recently, \citet{kacprzak19}
related~{\OVI} absorber kinematics to host galaxy rotation
curves. They found that along the projected galaxy major axis, where
accretion is expected, {\OVI} does not correlate with galaxy rotation
kinematics like {\MgII} \citep[e.g.,][]{steidel02,
  ggk-sims,kacprzak11kin, ho17}. For gas observed along the projected
galaxy minor axis, {\OVI} absorbers best match models of decelerating
outflows. Combined with simulations, the authors suggest that {\OVI}
is not an ideal probe of gas accretion or outflows, but rather traces
the virial temperature of the host halo.

The work presented here will address both the halo mass dependence of
{\OVI} absorber kinematics, and how {\OVI} gas flows relative to the
host galaxies by examining the absorber--galaxy kinematics, using a
subset of {\OVI} absorbers from the ``Multiphase Galaxy Halos''
Survey. We employ two TPCF methods: (1) absorber kinematics, which is
the approach employed by \citet{nielsen17}, and (2) absorber--galaxy
kinematics. In constructing the TPCFs for absorber--galaxy kinematics
(method 2), we apply the velocity offset between the absorber redshift
and the galaxy redshift. We also normalize the absorber--galaxy
velocities with respect to the circular velocity at the observed
impact parameter, $V_{\rm c}(D)$, to take into consideration the range
of halo masses in the sample \citep[similar to the normalization done
  in][]{magiicat4}. Average absorption profiles are presented to
complement the TPCFs by providing information about the optical depth.

In Section~\ref{sec:sample}, we present the sample and elaborate on
how the kinematics are quantified, namely with the TPCFs and average
absorption profiles. In Section~\ref{sec:bivariate-abs}, we present
the mass dependence of absorber kinematics, comparing our sample to
the group environment sample published in \citet{pointon17} and the
simulated aperture column densities presented by
\citet{oppenheimer16}. Section~\ref{sec:abs-galkin} presents new
absorber--galaxy kinematics for various subsamples segregated by
galaxy redshift, $z_{\rm gal}$, inclination, $i$, azimuthal angle,
$\Phi$, and halo mass, {\logtenmv}. In Section \ref{sec:discussion},
we discuss the halo mass dependence of {\OVI} absorber kinematics and
non-virialized motions in the form of outflows. Finally, we conclude
in Section \ref{sec:conclusions}.  Throughout we assume a $\Lambda$CDM
cosmology ($H_0 = 70{\rm\,km\,s^{-1}\,Mpc^{-1}}, \Omega_M=0.3,
\Omega_\Lambda=0.7$).

%%%%%%%%%%%%%%%%%%%%%%%%%%%%%%%%%%%%%%%%%%%%%%%%%%%%%%%%%%%%%%%%%%%%%%%%%%%%%%%
\begin{deluxetable*}{llcccccccccc}
	\tabletypesize{\footnotesize}
	\tablecolumns{12}
	\tablewidth{0pt}
	\setlength{\tabcolsep}{0.06in}
	\tablecaption{~{\OVI} Absorber--Galaxy Properties \label{tab:props}}
	\tablehead{
		\colhead{(1)}                      &
		\colhead{(2)}                      &
		\colhead{(3)}                      &
		\colhead{(4)}                      &
		\colhead{(5)}                      &
		\colhead{(6)}                      &
		\colhead{(7)}                      &
		\colhead{(8)}                      &
		\colhead{(9)}                      &
		\colhead{(10)}                     &
		\colhead{(11)}                     &
		\colhead{(12)}                     \\
		\colhead{Field}                    &
		\colhead{$z_{\rm gal}$}             &
		\colhead{Ref~\tablenotemark{a}}    &
		\colhead{$z_{\rm abs}$}             &
		\colhead{$W_{\rm r}(1031)$}         &
		\colhead{$\Phi$}                   &
		\colhead{$i$}                      &
		\colhead{$M_r$\tablenotemark{b}}   &
		\colhead{\logtenmv}                &
%		\colhead{$V_{\rm c}^{\rm max}$}    &
		\colhead{$V_{\rm c}(D)$}            &
		\colhead{$D$}                      &
		\colhead{$R_{\rm vir}$}             \\
		\cline{2-3}
		\colhead{}                         &
		\colhead{}                         &
		\colhead{}                         &
		\colhead{}                         &
		\colhead{(\AA)}                    &
		\colhead{(deg)}                    &
		\colhead{(deg)}                    &
		\colhead{(AB)}                     &
		\colhead{}                         &
%		\colhead{($\rm{km\, s^{-1}}$)}     &
		\colhead{($\rm{km\, s^{-1}}$)}     &		
		\colhead{(\rm{kpc})}               &
		\colhead{(\rm{kpc})}
	}
	\startdata
	${\rm J}012528-000555$  & $0.398525$ & 2  & $0.399090$ & $0.817$ & $73.4_{-4.7}^{+4.6}$   & $63.2_{-2.6}^{+1.7}$   & -21.99 & $12.51_{-0.15}^{+0.16}$ & $242.9_{-53.6}^{+68.8}$  & $163.0\pm0.1$ & $285.4_{-32.0}^{+37.3}$ \\[3.0pt]
${\rm J}035128-142908$  & $0.356992$ & 1  & $0.356825$ & $0.396$ & $4.9_{-40.2}^{+33.0}$  & $28.5_{-12.5}^{+19.8}$ & -20.86 & $12.00_{-0.19}^{+0.29}$ & $174.8_{-50.6}^{+90.9}$  & $72.3\pm0.4$  & $190.8_{-25.9}^{+47.9}$ \\[3.0pt]
${\rm J}040748-121136$  & $0.3422$   & 3  & $0.342042$ & $0.056$ & $48.1_{-0.9}^{+1.0}$   & $85.0_{-0.4}^{+0.1}$   & -19.77 & $11.62_{-0.21}^{+0.42}$ & $107.5_{-36.9}^{+83.8}$  & $172.0\pm0.1$ & $142.5_{-21.6}^{+53.9}$ \\[3.0pt]
${\rm J}040748-121136$  & $0.495164$ & 4  & $0.495101$ & $0.229$ & $21.0_{-3.7}^{+5.3}$   & $67.2_{-7.5}^{+7.6}$   & -19.73 & $11.41_{-0.21}^{+0.45}$ & $97.5_{-33.7}^{+82.6}$   & $107.6\pm0.4$ & $124.4_{-18.2}^{+51.6}$ \\[3.0pt]
${\rm J}045608-215909$  & $0.381511$ & 1  & $0.381514$ & $0.219$ & $63.8_{-2.7}^{+4.3}$   & $57.1_{-2.4}^{+19.9}$  & -20.87 & $12.00_{-0.19}^{+0.29}$ & $167.7_{-48.4}^{+86.8}$  & $103.4\pm0.3$ & $192.4_{-26.0}^{+48.0}$ \\[3.0pt]
${\rm J}091440+282330$  & $0.244312$ & 1  & $0.244098$ & $0.333$ & $18.2_{-1.0}^{+1.1}$   & $39.0_{-0.2}^{+0.4}$   & -20.55 & $11.88_{-0.20}^{+0.33}$ & $153.2_{-46.9}^{+91.5}$  & $105.9\pm0.1$ & $170.7_{-23.9}^{+49.4}$ \\[3.0pt]
${\rm J}094331+053131$  & $0.353052$ & 1  & $0.353286$ & $0.220$ & $8.2_{-5.0}^{+3.0}$    & $44.4_{-1.2}^{+1.1}$   & -19.88 & $11.66_{-0.21}^{+0.41}$ & $125.9_{-42.6}^{+95.4}$  & $96.5\pm0.3$  & $146.8_{-22.0}^{+54.1}$ \\[3.0pt]
${\rm J}094331+053131$  & $0.548494$ & 1  & $0.548769$ & $0.275$ & $67.2_{-1.0}^{+0.9}$   & $58.8_{-1.1}^{+0.6}$   & -21.30 & $11.96_{-0.18}^{+0.26}$ & $150.3_{-41.3}^{+70.2}$  & $150.9\pm0.6$ & $190.9_{-24.9}^{+42.9}$ \\[3.0pt]
${\rm J}095000+483129$  & $0.211866$ & 1  & $0.211757$ & $0.211$ & $16.6_{-0.1}^{+0.1}$   & $47.7_{-0.1}^{+0.1}$   & -21.73 & $12.37_{-0.16}^{+0.18}$ & $237.0_{-55.3}^{+74.7}$  & $93.6\pm0.2$  & $246.9_{-29.0}^{+36.1}$ \\[3.0pt]
${\rm J}100402+285535$  & $0.1380$   & 5  & $0.137724$ & $0.117$ & $12.4_{-2.9}^{+2.4}$   & $79.1_{-2.1}^{+2.2}$   & -17.05 & $10.87_{-0.22}^{+0.63}$ & $69.9_{-27.6}^{+89.3}$   & $56.7\pm0.2$  & $76.3_{-11.7}^{+47.3}$ \\[3.0pt]
${\rm J}100902+071343$  & $0.227855$ & 1  & $0.227851$ & $0.576$ & $89.6_{-1.3}^{+1.3}$   & $66.3_{-0.9}^{+0.6}$   & -20.19 & $11.76_{-0.21}^{+0.37}$ & $149.0_{-48.2}^{+101.2}$ & $64.0\pm0.8$  & $154.5_{-22.5}^{+50.9}$ \\[3.0pt]
${\rm J}104116+061016$  & $0.442173$ & 1  & $0.441630$ & $0.368$ & $4.3_{-1.0}^{+0.9}$    & $49.8_{-5.2}^{+7.4}$   & -21.36 & $11.99_{-0.18}^{+0.26}$ & $173.8_{-47.1}^{+79.1}$  & $56.2\pm0.3$  & $193.1_{-24.9}^{+42.2}$ \\[3.0pt]
${\rm J}111908+211918$  & $0.1383$   & 6  & $0.138521$ & $0.074$ & $34.4_{-0.4}^{+0.4}$   & $26.4_{-0.4}^{+0.8}$   & -21.45 & $12.24_{-0.17}^{+0.21}$ & $204.3_{-51.4}^{+76.5}$  & $138.0\pm0.2$ & $219.0_{-27.1}^{+38.8}$ \\[3.0pt]
${\rm J}113327+032719$  & $0.154599$ & 4  & $0.153979$ & $0.252$ & $56.1_{-1.3}^{+1.7}$   & $23.5_{-0.2}^{+0.4}$   & -19.84 & $11.64_{-0.21}^{+0.41}$ & $139.8_{-47.5}^{+107.1}$ & $55.6\pm0.1$  & $138.7_{-20.9}^{+51.6}$ \\[3.0pt]
${\rm J}113910-135043$  & $0.204194$ & 1  & $0.204297$ & $0.231$ & $5.8_{-0.5}^{+0.4}$    & $83.4_{-0.5}^{+0.4}$   & -19.99 & $11.69_{-0.21}^{+0.40}$ & $133.0_{-44.3}^{+98.0}$  & $93.2\pm0.3$  & $146.2_{-21.6}^{+52.4}$ \\[3.0pt]
${\rm J}113910-135043$  & $0.212259$ & 1  & $0.212237$ & $0.137$ & $80.4_{-0.5}^{+0.4}$   & $85.0_{-0.6}^{+5.0}$   & -20.09 & $11.73_{-0.21}^{+0.39}$ & $119.2_{-39.2}^{+84.4}$  & $174.8\pm0.1$ & $150.3_{-22.0}^{+51.7}$ \\[3.0pt]
${\rm J}113910-135043$  & $0.219724$ & 4  & $0.219820$ & $0.021$ & $44.9_{-8.1}^{+8.9}$   & $85.0_{-8.5}^{+5.0}$   & -17.67 & $11.04_{-0.21}^{+0.60}$ & $66.8_{-25.8}^{+80.5}$   & $122.0\pm0.2$ & $88.7_{-13.5}^{+52.0}$  \\[3.0pt]
${\rm J}113910-135043$  & $0.319255$ & 1  & $0.319167$ & $0.255$ & $39.1_{-1.7}^{+1.9}$   & $83.4_{-1.1}^{+1.4}$   & -20.48 & $11.86_{-0.20}^{+0.34}$ & $157.0_{-48.5}^{+96.1}$  & $73.3\pm0.4$  & $170.4_{-23.9}^{+50.7}$ \\[3.0pt]
${\rm J}121920+063838$  & $0.1241$   & 6  & $0.124103$ & $0.424$ & $67.2_{-91.4}^{+39.8}$ & $22.0_{-21.8}^{+18.7}$ & -20.50 & $11.87_{-0.20}^{+0.34}$ & $156.8_{-48.4}^{+95.6}$  & $93.4\pm5.3$  & $163.1_{-22.9}^{+48.2}$ \\[3.0pt]
${\rm J}123304-003134$  & $0.318757$ & 4  & $0.318609$ & $0.439$ & $17.0_{-2.3}^{+2.0}$   & $38.7_{-1.8}^{+1.6}$   & -20.62 & $11.91_{-0.20}^{+0.32}$ & $159.2_{-48.3}^{+92.5}$  & $88.9\pm0.2$  & $176.5_{-24.7}^{+49.7}$ \\[3.0pt]
${\rm J}124154+572107$  & $0.205267$ & 1  & $0.205538$ & $0.519$ & $77.6_{-0.4}^{+0.3}$   & $56.4_{-0.5}^{+0.3}$   & -19.83 & $11.64_{-0.21}^{+0.41}$ & $145.0_{-49.3}^{+114.4}$ & $21.1\pm0.1$  & $140.2_{-21.1}^{+52.3}$ \\[3.0pt]
${\rm J}124154+572107$  & $0.217905$ & 4  & $0.218043$ & $0.366$ & $63.0_{-2.1}^{+1.8}$   & $17.4_{-1.6}^{+1.4}$   & -19.77 & $11.62_{-0.21}^{+0.42}$ & $124.0_{-42.5}^{+96.7}$  & $94.6\pm0.2$  & $138.7_{-21.0}^{+52.4}$ \\[3.0pt]
${\rm J}124410+172104$  & $0.5504$   & 5  & $0.550622$ & $0.447$ & $20.1_{-19.1}^{+16.7}$ & $31.7_{-4.8}^{+16.2}$  & -20.97 & $11.82_{-0.19}^{+0.31}$ & $144.4_{-42.3}^{+79.0}$  & $21.2\pm0.3$  & $171.2_{-23.1}^{+45.5}$ \\[3.0pt]
${\rm J}131956+272808$  & $0.6610$   & 7  & $0.660670$ & $0.311$ & $86.6_{-1.2}^{+1.5}$   & $65.8_{-1.2}^{+1.2}$   & -21.70 & $12.15_{-0.15}^{+0.19}$ & $183.6_{-41.6}^{+59.8}$  & $103.9\pm0.5$ & $223.9_{-24.9}^{+34.7}$ \\[3.0pt]
${\rm J}132222+464546$  & $0.214431$ & 1  & $0.214320$ & $0.354$ & $13.9_{-0.2}^{+0.2}$   & $57.9_{-0.2}^{+0.1}$   & -21.18 & $12.13_{-0.18}^{+0.25}$ & $204.0_{-54.7}^{+90.8}$  & $38.6\pm0.2$  & $204.6_{-26.1}^{+43.8}$ \\[3.0pt]
${\rm J}134251-005345$  & $0.227042$ & 1  & $0.227196$ & $0.373$ & $13.2_{-0.4}^{+0.5}$   & $0.1_{-0.1}^{+0.6}$    & -21.77 & $12.39_{-0.16}^{+0.17}$ & $239.3_{-55.2}^{+73.7}$  & $35.3\pm0.2$  & $251.7_{-29.3}^{+35.9}$ \\[3.0pt]
${\rm J}155504+362847$  & $0.189201$ & 1  & $0.189033$ & $0.385$ & $47.0_{-0.8}^{+0.3}$   & $51.8_{-0.7}^{+0.7}$   & -21.03 & $12.07_{-0.18}^{+0.27}$ & $196.3_{-54.2}^{+93.8}$  & $33.4\pm0.1$  & $193.7_{-25.2}^{+44.6}$ \\[3.0pt]
${\rm J}213135-120704$  & $0.4302$   & 8  & $0.430164$ & $0.385$ & $14.9_{-4.9}^{+6.0}$   & $48.3_{-3.7}^{+3.5}$   & -21.47 & $12.04_{-0.18}^{+0.25}$ & $181.3_{-48.1}^{+79.0}$  & $48.4\pm0.2$  & $199.7_{-25.3}^{+41.8}$ \\[3.0pt]
${\rm J}225357+160853$  & $0.153718$ & 9  & $0.153821$ & $0.263$ & $59.6_{-1.8}^{+0.9}$   & $33.3_{-2.0}^{+2.7}$   & -19.55 & $11.55_{-0.21}^{+0.45}$ & $137.8_{-48.1}^{+115.4}$ & $31.8\pm0.2$  & $129.5_{-19.6}^{+52.7}$ \\[3.0pt]
${\rm J}225357+160853$  & $0.352787$ & 1  & $0.352708$ & $0.381$ & $88.7_{-4.8}^{+4.6}$   & $36.7_{-4.6}^{+6.9}$   & -20.67 & $11.93_{-0.20}^{+0.32}$ & $138.1_{-41.6}^{+78.4}$  & $203.2\pm0.5$ & $180.3_{-25.2}^{+49.5}$ \\[3.0pt]
${\rm J}225357+160853$  & $0.390013$ & 9  & $0.390705$ & $0.173$ & $24.2_{-1.2}^{+1.2}$   & $76.1_{-1.2}^{+1.1}$   & -21.25 & $12.16_{-0.18}^{+0.24}$ & $160.4_{-42.5}^{+69.0}$  & $276.3\pm0.2$ & $217.2_{-27.5}^{+44.9}$ \\[3.0pt]

%$221.8_{-47.3}^{+61.1}$ & 
%$150.0_{-42.2}^{+75.7}$ &
%$112.5_{-37.6}^{+85.0}$ & 
%$94.5_{-31.9}^{+77.6}$  & 
%$150.1_{-42.1}^{+75.3}$ & 
%$139.0_{-41.4}^{+80.5}$ & 
%$115.5_{-38.1}^{+84.9}$ & 
%$143.7_{-38.3}^{+65.1}$ & 
%$203.3_{-45.9}^{+62.3}$ & 
%$64.8_{-25.1}^{+80.4}$  & 
%$126.5_{-39.9}^{+83.4}$ & 
%$148.0_{-39.1}^{+65.4}$ & 
%$186.0_{-45.4}^{+67.7}$ & 
%$117.0_{-38.8}^{+86.9}$ & 
%$120.7_{-39.2}^{+86.3}$ & 
%$123.7_{-39.6}^{+85.0}$ & 
%$72.8_{-27.6}^{+85.1}$  & 
%$135.4_{-40.7}^{+80.4}$ & 
%$139.4_{-41.7}^{+82.3}$ &
%$140.3_{-41.4}^{+79.1}$ & 
%$115.8_{-38.4}^{+86.3}$ & 
%$114.0_{-38.1}^{+86.2}$ & 
%$128.7_{-36.7}^{+68.3}$ & 
%$165.1_{-36.3}^{+52.2}$ & 
%$168.5_{-43.9}^{+72.8}$ & 
%$206.2_{-46.0}^{+61.8}$ & 
%$160.9_{-43.2}^{+74.6}$ & 
%$154.1_{-39.7}^{+65.2}$ & 
%$109.2_{-37.2}^{+88.7}$ & 
%$141.9_{-41.6}^{+78.2}$ & 
%$169.2_{-43.5}^{+70.7}$ & 

	\enddata

\tablenotetext{a}{Galaxy redshift references: (1) \citet{kacprzak19},
  (2) \citet{muzahid15}, (3) \citet{johnson13}, (4) \citet{pointon19},
  (5) \citet{chen01a}, (6) \citet{prochaska11}, (7) \citet{ggk1317},
  (8) \citet{guillemin97}, and (9) this work.}

\tablenotetext{b}{Galaxy absolute $r$-band magnitude in the AB
  system. The magnitudes are converted into Vega mags with $M_r-5\log
  h=M_r(AB)-5\log(0.7)-0.1429$, which are then used in the halo
  abundance matching method \citep[for details, see][]{magiicat3}.}

\end{deluxetable*}
%%%%%%%%%%%%%%%%%%%%%%%%%%%%%%%%%%%%%%%%%%%%%%%%%%%%%%%%%%%%%%%%%%%%%%%%%%%%%%%

\section{Sample and Data Analysis}
\label{sec:sample}

The sample of {\OVI} absorber--galaxy pairs used in this work is a
subset of the ``Multiphase Galaxy Halos'' Survey \citep{kacprzak15,
  kacprzak19, muzahid15, muzahid16, nielsen17, pointon17,
  pointon19}. The associated galaxy spectroscopic redshifts, $z_{\rm
  gal}$, spanning $0.1241 \leq z_{\rm gal} \leq 0.6610$ (median
$\braket{z_{\rm gal}} = 0.2443$), are accurate to within $\sigma_z
\leq 0.0001$, which is $\sim30\rm{\,km\,s^{-1}}$ in velocity space
\citep[e.g.,][]{kacprzak19}. The absorber--galaxy pairs are also
within an on-the-sky projected distance (impact parameter) of $D
\approx 200$~kpc (21.1~kpc$< D < 276.3$~kpc, median $\braket{D} =
93.2$~kpc). All of the galaxies are isolated, that is, there were no
identified neighboring galaxies within a projected distance of 200~kpc
from the line-of-sight of the quasar, and within a line-of-sight
velocity separation of 500~{\kms}. Absorption systems with
line-of-sight velocities larger than $\pm500$~{\kms} away from their
identified host galaxies are assumed not to be associated with the
host galaxy. The sample is somewhat heterogeneous in that the quasar
fields were drawn from several works \citep[see Table~\ref{tab:props}
  and][]{kacprzak19} and each survey has different completeness
levels. However, the quasar fields have generally been surveyed to a
sensitivity of $0.1L^{\ast}$ out to at least 350~kpc, with the
exception of galaxies from the COS-Halos survey, of which there are
12, which goes out to only 150~kpc \citep[for details,
  see][]{tumlinson13, werk13}.

For our {\OVI} absorber kinematics analysis
(Section~\ref{sec:bivariate-abs}), we also include a sample of six
galaxy group environments from \cite{pointon17} for comparison. See
\cite{pointon17} for further details. We include this sample to cover
a large range in halo masses to investigate the {\OVI} column density
dependence on halo mass found in the EAGLE simulations
\citep{oppenheimer16}.

\subsection{Galaxy Properties}

We have selected 31 absorber--galaxy pairs from the ``Multiphase
Galaxy Halos" Survey that are suitable for this study. Each galaxy in
the sample was imaged with ACS, WFC3, or WFPC2 on the {\it Hubble
  Space Telescope (HST)}. GIM2D \citep{simard02} was then used to
model the morphological properties of the galaxies; the details of the
modeling are elaborated upon in \citet{kacprzak15}. Following
\citet{nielsen17}, we define galaxies having inclination angles of
$0^\circ \leq i < 51^\circ$ as face-on, and galaxies with $51^\circ
\leq i \leq 90^\circ$ as edge-on. We also define azimuthal angles of
$0^\circ \leq \Phi < 45^\circ$ as a quasar sightline aligned with the
projected major axis of the galaxy, and azimuthal angles of $45^\circ
\leq \Phi \leq 90^\circ$ as a quasar sightline aligned with the
projected minor axis of the galaxy.

To measure accurate redshifts, galaxy spectra were obtained for a
majority of our sample (24/31) using the Keck Echellette Spectrograph
and Imager, ESI \citep{sheinis02}. Details of the observations and
data reduction, and most of the new redshifts are presented in
\citet{kacprzak19}. Several additional redshifts determined with this
method are presented in \citet{pointon19} and here. The ESI spectra
have a resolution of 22~\kms~pixel$^{-1}$ (${\rm FWHM}\sim90$~km/s)
when binned by two in the spectral direction and have a wavelength
coverage of $4000$ to $10000$~\AA, which allows for detection of
multiple emission lines such as the [{\OII}] doublet, $\rm{H}\beta$,
[{\OIII}] doublet, $\rm{H}\alpha$, and [\NII] doublet. Galaxy spectra
are both vacuum and heliocentric velocity-corrected to provide a
direct comparison with the absorption line spectra. The Gaussian
fitting algorithm {\sc fitter} \citep[][]{archiveI} was used to
compute best-fit emission-line centroids and widths to derive galaxy
redshifts. Galaxy redshifts obtained with ESI have accuracies ranging
from $3-20$~\kms. To test for possible systematic shifts in our
wavelength solutions that could increase our uncertainties, the
derived wavelength solution was verified against a catalog of known
sky lines ranging between $4000-10000$~{\AA}. This test resulted in an
rms difference of $\sim0.03$~{\AA} ($\sim2$~{\kms}), which is lower
than the galaxy redshift error obtained by fitting multiple emission
and absorption lines. The remainder of the galaxy redshifts were
obtained from previous studies, and are tabulated in Table
\ref{tab:props} \citep{guillemin97, chen01a, prochaska11, ggk1317,
  johnson13}.

Since previous work has suggested that halo mass is key to governing
the presence or absence of {\OVI} absorbing gas
\citep[e.g.,][]{oppenheimer16, pointon17}, we calculate halo masses,
circular velocities, and virial radii for each of the galaxies in our
sample to investigate this mass dependence for gas kinematics. We
follow the halo abundance matching method described in Appendix A of
\citet{magiicat3} and summarize it here \citep[also
  see][]{trujillo11}. Halo abundance matching makes the assumption
that the number density of galaxies with a given observed galaxy
property \citep[in this case, the COMBO-17 $r-$band luminosity
  functions from][]{wolf03} is mapped to the distribution function of
simulated galaxies with a given property \citep[the maximum circular
  velocity or dark matter halo mass from the Bolshoi $N$-body
  cosmological simulation dark matter halo catalogs;][]{klypin11}. For
each galaxy in our sample, the $r-$band absolute Vega magnitude,
$M_r-5\log({\rm h})$, was calculated and used in the corresponding redshift
curves from Figure~9(b) of \citet{magiicat3} to calculate halo masses,
${\logtenmv}$.~From the halo masses, we calculate the virial radii
according to \citet{bryan98}. Circular velocities, $V_{\rm c}(D)$,
were calculated at the impact parameter of absorption using Equations
5 of \citet{nfw96} and B2 of \citet{magiicat3}. In the sample, the
halo masses span $10.87 \leq {\logtenmv} \leq 12.51$ (median
$\braket{\logtenmv} = 11.88$); the circular velocities span
$66.8\ \kms \leq V_{\rm c}(D) \leq 242.9\ \kms$ (median
$\braket{V_{\rm c}(D)} = 150.3$~{\kms}).

In Table~\ref{tab:props}, we list the absorber--galaxy pairs used in
this work, the corresponding galaxy redshifts, $z_{\rm gal}$, {\OVI}
absorber redshifts, $z_{\rm abs}$, rest-frame equivalent widths,
$W_{\rm r}(1031)$, azimuthal angle, $\Phi$, galaxy inclination, $i$,
absolute $r$-band (AB) magnitude, $M_r$, halo mass, ${\logtenmv}$,
circular velocity at the observed impact parameter, $V_{\rm c}(D)$,
impact parameter, $D$, and the virial radius, $R_{\rm vir}$. The
subsample cuts, number of galaxies, median halo masses, and median
redshifts for each subsample are listed in
Table~\ref{tab:tpcfprops_OVI}.

In the sample, rank correlation tests yield no statistically
significant correlations between the galaxy redshift, azimuthal angle,
inclination, and halo mass. A one-dimensional Kolmogorov-Smirnov (KS)
test was also carried out on the galaxy orientation measurements to
test whether the sample is unbiased. We find that the azimuthal angles
of the galaxies are consistent with that of unbiased samples at the
$0.6\sigma$ level; the inclination angles of the galaxies are also
consistent with that of unbiased samples at the $2.3\sigma$ level.

\subsection{Quasar Spectra}

The details of the quasar spectra are found in \citet{kacprzak15} and
\citet{nielsen17}, but we summarize them here. Each of the 23 quasars
has a medium resolution ($R\sim20,000,\ \rm{FWHM}\sim18$~{\kms})
spectrum from \emph{HST}/COS. Voigt profiles were fitted to each of
the {\OVI} $\lambda\lambda1031,1037$ doublet absorption lines with
\textsc{VPFIT} \citep{carswell14}, and the zero-points of velocity
(i.e., $z_{\rm abs}$) were defined as the velocity where $50\%$ of the
modeled absorption resides on each side in the optical depth
distribution for the {\OVI} $\lambda 1031$ line \citep{nielsen17}. The
velocity bounds of the absorption were defined to be where modeled
absorption deviates from the continuum (value of 1) by $1\%$ (to
0.99). The absorption profiles for each absorber--galaxy pair are
plotted in Appendix~\ref{app:spectra}, where the velocity zero-point
corresponds to the systemic velocity of the host galaxy.

\subsection{Pixel-Velocity Two-Point Correlation Function}

The TPCF method has previously been used to analyze the absorber
velocity dispersions of {\MgII} and {\OVI} absorbers surrounding
galaxies \citep{magiicat5, magiicat4, nielsen17, magiicat6,
  pointon17}. In the first part of this work, we investigate the mass
dependence of the velocity dispersions of the absorbers (``absorber
kinematics''). For the rest of this work, we shift the velocities
relative to the galaxy systemic velocity to investigate the motion of
the surrounding gas relative to the galaxy (``absorber--galaxy
kinematics'').

\subsubsection{Absorber Kinematics}

The details of the absorber TPCF construction are expounded in
\citet{magiicat4}, but we briefly summarize the method here. The
velocities of all the pixels, within the velocity bounds in which
{\OVI} absorption is formally detected, are first extracted for a
desired subsample (e.g., all edge-on galaxies) and combined into a
single array. Statistically, for a given subsample, this step makes
the equivalency between a single quasar absorption sightline around
multiple galaxies, and a single galaxy with multiple sightlines. We
then calculate the absolute value of the velocity separations between
every pair of pixel velocities in the subsample, $\Delta v_{\rm
  pix}$. These velocity separations are binned into 20~{\kms} bins to
account for the resolution of COS, and the number of counts in each bin
is normalized by the total number of velocity separation pairs in the
subsample. This yields a probability distribution function of the
velocity dispersion, the absorber TPCF.

The uncertainties on the TPCFs were determined by a bootstrap analysis
with 100 bootstrap realizations where absorber--galaxy pairs were
randomly drawn with replacement. These are $1\sigma$ uncertainties
from the mean of the bootstrap realizations. To characterize the
TPCFs, we use the quantities $\Delta v(50)$ and $\Delta v(90)$. These
represent the velocity separations within which $50\%$ and $90\%$ of
the area under the TPCF is located. The corresponding uncertainties
for $\Delta v(50)$ and $\Delta v(90)$ were also determined from the
bootstrap analysis, where we obtained the $1\sigma$ uncertainties from
the mean of the bootstrap realizations. Following \cite{magiicat5}, we
employ two-sample $\chi^2$ tests to examine and quantify statistical
differences between the TPCFs of different galaxy-absorber subsamples.
We report the reduced $\chi^2$, i.e., $\chi^2_\nu$, where $\nu$ is the
number of degrees of freedom.

\subsubsection{Absorber--Galaxy Kinematics}
\label{sec:TPCFabs-gal}

The TPCFs described in the previous section were modified to account
for the velocity of the gas relative to the host galaxy. After the
pixel velocities are extracted from a subsample, they are shifted with
respect to the host galaxy,
\begin{equation}
v_{\rm pix-gal} = \left|v_{\rm pix} + c\frac{z_{\rm abs} - z_{\rm gal}}{1 + z_{\rm gal}}\right|.
\end{equation}
We take the absolute value of this shifted velocity to quantify the
velocity dispersion of the absorbing gas with respect to the galaxy
systemic velocity, without considering its direction. We also do not
know whether the gas is physically located in front of or behind the
host galaxy, so we cannot determine if the gas is infalling or
outflowing relative to the galaxy, hence the velocity sign is not
important. Thus, in this work, the absorber--galaxy TPCF is defined to
be a statistical measure of the velocity dispersion of the absorbers
whose velocities are shifted with respect to the galaxy systemic
velocity.

Since our sample of galaxies spans a range of halo masses, we account
for the galaxy halo mass by normalizing the shifted velocities by the
circular velocity at the observed impact parameter of the host galaxy,
$V_{\rm c}(D)$. We now work with the circular velocity-normalized
pixel--galaxy velocities, $v_{\rm pix-gal}/V_{\rm c}(D)$. Once these
values are obtained, the normalized velocities for a given subsample
are combined and are subtracted between every possible pair of
pixels. Thus we obtain normalized pixel--galaxy velocity separations,
$\Delta (v_{\rm pix-gal}/V_{\rm c}(D))$.

For the absorber--galaxy TPCF, we use a bin size of $\Delta(v_{\rm
  pix-gal}/V_{\rm c}(D)) = 0.2$, which is determined by dividing the
maximum uncertainty of the systemic galaxy redshift, $\Delta z_{\rm
  gal} = 0.0001$, corresponding to $\sim30/(1+\braket{z_{\rm
    gal}})$~{\kms} in the galaxy rest frame, by the average $V_{\rm
  c}(D)$ in the sample, $156$~{\kms}.

\subsection{Average Absorption Profiles}

We complement the TPCFs with the average absorption profiles for a
given subsample. The average profiles provide supplementary
information about the optical depth distribution and indicate how
much gas there is at a given velocity. 

The average absorption profiles were constructed by first extracting
the pixel velocities from the individual {\OVI} $\lambda1031$
absorption profiles. For absorber kinematics, we do not modify these
velocities. For absorber--galaxy kinematics, we shift the velocities
relative to the galaxy systemic velocity and normalize them with
respect to the galaxy's circular velocity, $V_{\rm c}(D)$. The
associated flux values, obtained using model absorption profiles to
the data to remove any contamination due to noise and blends on the
flux, are also extracted. Next, for any given profile, the positive
pixel velocities and negative pixel velocities are separated into two
arrays. We take the absolute value of the negative velocities because
we do not know where the gas is located relative to the galaxy other
than the projected distance, thus the sign on the velocities bears
little meaning. Then for both the velocity and flux arrays we run a
linear interpolation routine onto a standardized velocity grid to
place all absorbers on the same velocity or absorber--galaxy
normalized velocity scale. This is repeated for all the
absorber--galaxy pairs in the subsample.

The final average absorption profile is then obtained by calculating
the average flux over all absorber--galaxy pairs in the subsample, for
each velocity bin. The uncertainties on the absorption, like the
TPCFs, were determined by a bootstrap analysis with 100 bootstrap
realizations, where absorber--galaxy pairs were randomly drawn with
replacement. These are $1\sigma$ uncertainties from the mean of the
bootstrap realizations.

The individual {\OVI} absorption profiles, shifted relative to the
galaxy systemic velocity, are plotted in Appendix~\ref{app:spectra}
for direct comparison to the absorber--galaxy kinematics. 
$ $ \\
$ $ \\

\section{Absorber Kinematics}
\label{sec:bivariate-abs}

\begin{figure}[t]
	\includegraphics[width=\linewidth]{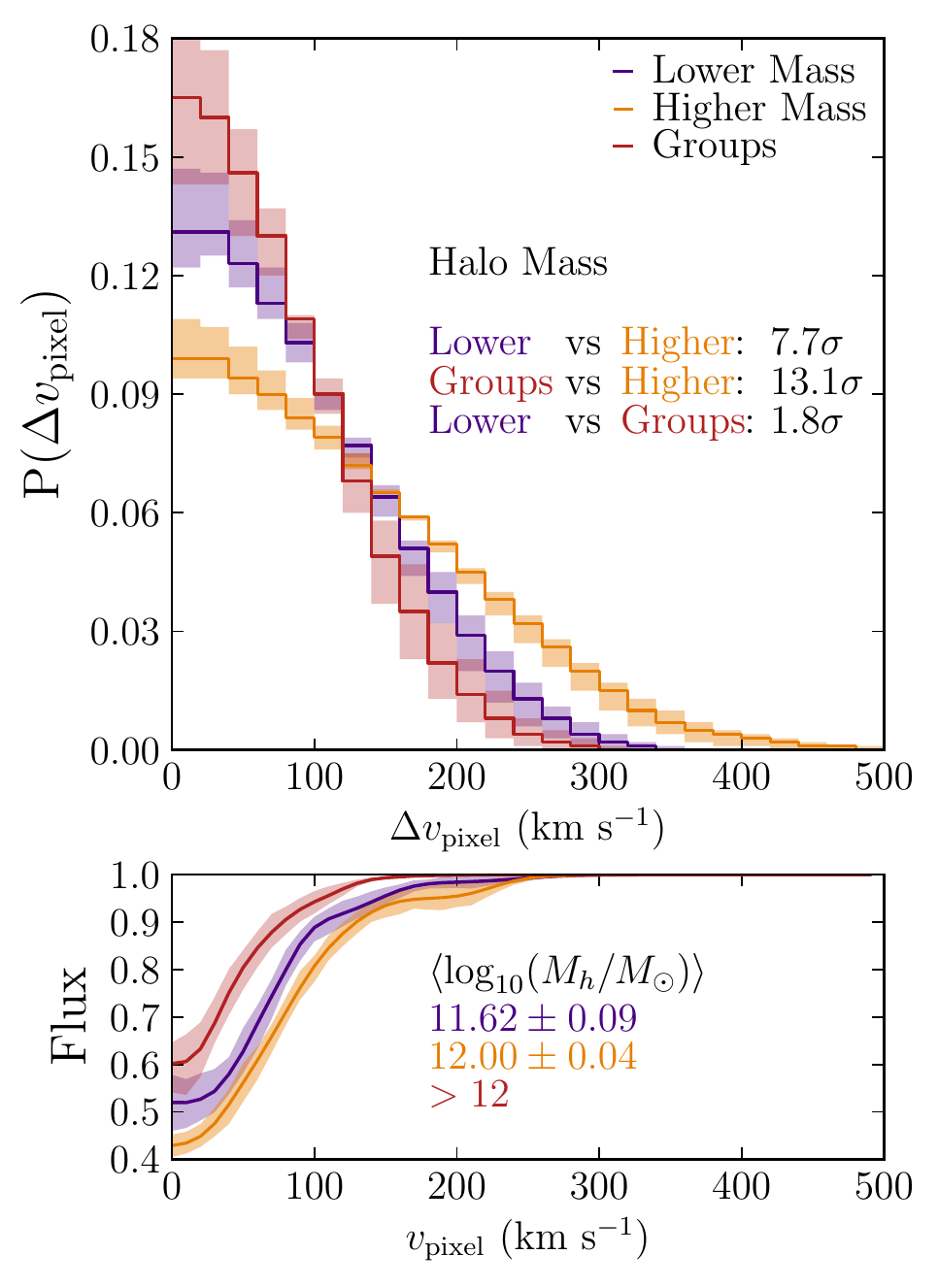}
	\caption[]{(Top) Absorber TPCFs for lower mass (purple),
          higher mass (orange), and group (red) galaxies. The higher
          mass galaxies have significantly larger velocity dispersions
          compared with the lower mass galaxies ($7.7\sigma$) and
          group galaxies ($13.1\sigma$). Group galaxies from
          \citet{pointon17} have similar velocity dispersions to lower
          mass galaxies. (Bottom) Average absorption profiles
          corresponding to lower mass galaxies, higher mass galaxies,
          and group galaxies. Group galaxies have lower optical depth
          at all pixel velocities compared with lower and higher mass
          galaxies.~$1\sigma$ uncertainties for both the TPCFs and
          average absorption spectra were calculated with a bootstrap
          analysis with 100 bootstrap realizations.}
	\label{fig:unnorm_masscutovi}
\end{figure}

\begin{figure}[t]
  \includegraphics[width=\linewidth]{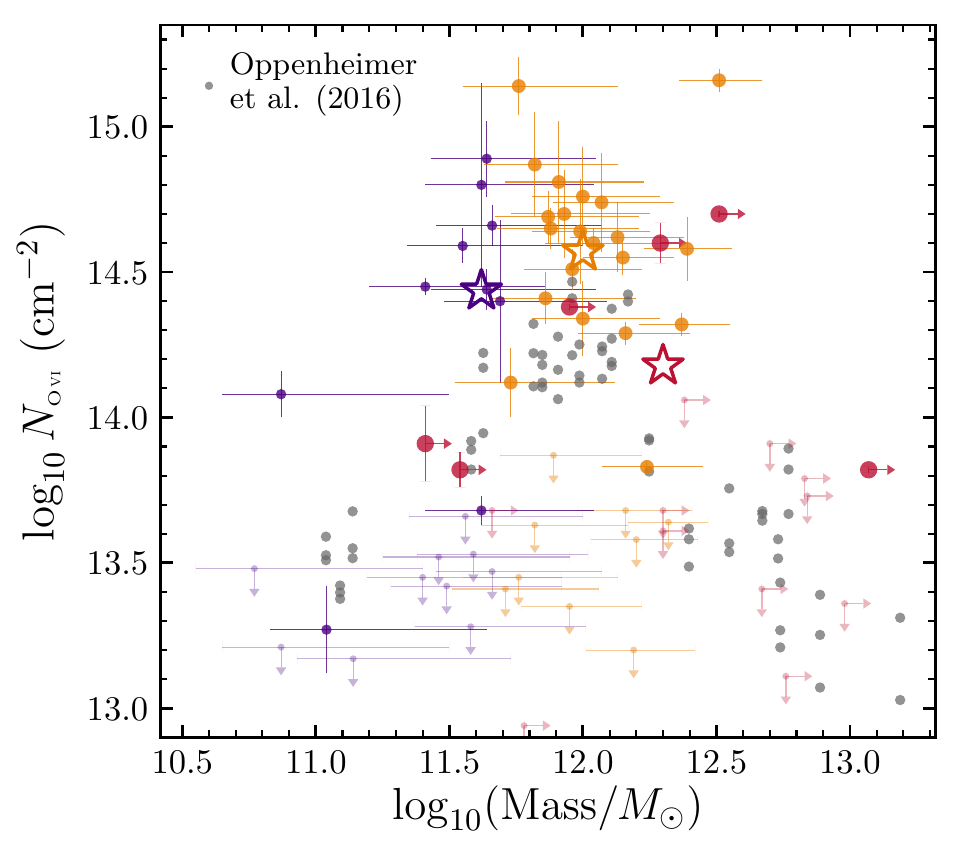}
  \caption[]{Column densities and masses for observational and
    simulation data. Lower mass galaxies are represented by purple
    points, higher mass are orange, and group galaxies are
    red. Because the group galaxy masses are difficult to estimate,
    their masses are plotted as lower limits. Upper limits on
    absorption are plotted as small points with downward arrows. Stars
    indicate the column densities obtained from the average absorption
    profiles presented in Figure~\ref{fig:unnorm_masscutovi}, where
    the halo mass is determined by the median halo mass for the lower
    and higher mass subsamples. Gray points are the aperture column
    densities within 150~kpc and $M_{200}$ masses from
    \citet{oppenheimer16}. Our data follow the trend of increasing
    column densities toward a maximum for $L^{\ast}$ galaxies, and
    decreasing toward the highest masses.}
  \label{fig:colopp}
\end{figure}

We first investigate the dependence of {\OVI} absorber kinematics on
halo mass in Figure~\ref{fig:unnorm_masscutovi}. The isolated galaxy
sample is sliced by {\logtenmv}$\ =11.7$, which was motivated by
\citet{oppenheimer16} who used it to define sub-$L^*$ and $L^*$
galaxies, which we call ``lower mass'' and ``higher mass'' galaxies
here, respectively. We also show the group galaxy sample from
\citet{pointon17} to study a more complete mass range for better
comparison with the simulations. The group sample was defined by the
number of galaxies and not mass since halo masses derived by halo
abundance matching may not be representative of the entire group
halo. \citet{oppenheimer16} defined group halos as those with
{\logtenmv}$\geq12.3$, but here we assume a more conservative mass of
{\logtenmv}$>12$ in the TPCF studies. If we calculate a mass for each
group member galaxy with halo abundance matching and sum these values,
we can obtain a lower limit on the group mass. We only use these
values in Figure~\ref{fig:colopp}. While these group masses span all
three mass subsamples, the group environments may have different
absorption characteristics due to interaction effects
\citep{alonso12b, fernandez15, pointon17} so we consider these to be a
separate sample.

In Figure~\ref{fig:unnorm_masscutovi}, higher mass galaxies have
significantly larger velocity dispersions than lower mass galaxies
($7.7\sigma$) and group galaxies ($13.1\sigma$), while group galaxies
have similar velocity dispersions to lower mass galaxies
($1.8\sigma$). From Table~\ref{tab:tpcfprops_OVI}, the $\Delta v(50)$
and $\Delta v(90)$ measurements for the lower mass galaxies
($80_{-8}^{+5}$~{\kms} and $188_{-19}^{+11}$~{\kms}, respectively) and
group galaxies ($64_{-7}^{+9}$~{\kms} and $153_{-18}^{+21}$~{\kms},
respectively) are all consistent within uncertainties. However, higher
mass galaxies have significantly larger $\Delta v(50)$
($108_{-9}^{+6}$~{\kms}) and $\Delta v(90)$ ($255_{-22}^{+14}$~{\kms})
values than for either lower mass galaxies or group galaxies.

We tested the robustness of this trend to outliers in column density
by redo-ing the TPCF analysis without the two absorber--galaxy pairs
corresponding to $\log N_{\tiny \OVI} > 15.0$. The significances
change slightly to $5.7\sigma, 12.2\sigma$, and $1.8\sigma$,
respectively, but the inferences remain the same. Additionally, we
tested whether the column density variation in the three different
mass bins was the cause for the kinematic trends, where total column
density and total velocity spread may be related. For this we
normalized each pixel velocity by the total column density for the
given absorber. While the results are not plotted here, we found that
the significances did not change drastically, with $8.0\sigma$,
$12\sigma$, and $0.7\sigma$, respectively. Thus the results we present
in Figure~\ref{fig:unnorm_masscutovi} are kinematic trends rather than
some underlying column density trend.

This effect is also observed in the average absorption profiles
plotted in the bottom panel of
Figure~\ref{fig:unnorm_masscutovi}. Higher mass galaxies have the most
optical depth at all line of sight velocities compared to lower mass
and group galaxies, while group galaxies have the least optical depth
at all line of sight velocities. We determined column densities for
each average absorption profile by mirroring the plotted profile over
$v_{\rm pixel}=0$~{\kms} (the resulting profile is symmetric) and
modeling the profiles with VPFIT as was done for the actual {\OVI}
profiles. We find column densities of $\log N_{\tiny \OVI}=14.570$ for
higher mass galaxies, $\log N_{\tiny \OVI}=14.44$ for lower mass
galaxies, and $\log N_{\tiny \OVI}=14.18$ for group galaxies. As
expected, the higher mass galaxies have the largest column densities
and group galaxies have the smallest.

Because we are probing different mass galaxies with our quasar
sightlines, we may be biased toward probing the inner regions of more
massive galaxies. This is important because {\OVI} equivalent widths
and column densities decrease with increasing impact parameter
\citep[e.g.,][]{tumlinson11, kacprzak15}. The median impact parameter
normalized by the virial radius, $D/R_{\rm vir}$, for lower mass
galaxies is 0.67, for higher mass galaxies is 0.46, and for group
galaxies is 0.47. However, a KS test comparing the $D/R_{\rm vir}$
distributions suggest that the lower and higher mass galaxy subsamples
were drawn from the same population ($2.1\sigma$). Comparing to the
group galaxy sample, we find significances of $0.0\sigma$ for group
galaxies versus higher mass galaxies and $0.6\sigma$ for group
galaxies versus lower mass galaxies. The $D/R_{\rm vir}$ ranges for
all three subsamples are also consistent. Thus differences in
$D/R_{\rm vir}$ between the subsamples does not appear to strongly
influence our results.

To better compare the kinematics results with the
\citet{oppenheimer16} simulations, we plot our data and the simulation
data on the column density--mass plane in Figure~\ref{fig:colopp}. The
observational data plotted are line-of-sight column densities and halo
masses from halo abundance matching. Halo masses of group galaxies are
measured using the same method, but the individual galaxy masses are
summed for each group to give a lower limit. Points are colored and
sized by the three halo mass bins from
Figure~\ref{fig:unnorm_masscutovi}. The simulation data are plotted as
gray points and represent aperture column densities within $D=150$~kpc
as a function of $\log(M_{200}/M_{\odot})$. The column densities
measured from the average absorption profiles in the previous
paragraph are plotted as stars for each mass subsample. Our data
generally follow the simulated trend that column densities increase
with increasing mass up to $\logtenmv\sim12.0$ and decrease with mass
above. Note that upper limits on the {\OVI} column densities from
\citet{kacprzak15} and \citet{pointon17} are plotted for completeness,
but they are not studied here since we focus on kinematics, which
non-absorbers do not have by definition. The upper limits mostly lie
below $\log N_{\tiny \OVI}=13.7$ regardless of halo mass, indicating
that the {\OVI} CGM is inherently patchy. We leave further analysis to
later work.

\renewcommand{\arraystretch}{1.3}
\begin{deluxetable*}{lccccccc}
	\tabletypesize{\footnotesize}
	\tablecolumns{8}
	\tablewidth{0pt}
	\setlength{\tabcolsep}{0.1cm}
	\tablecaption{TPCF $\Delta v(50)$ and $\Delta v(90)$ measurements \label{tab:tpcfprops_OVI}}
	\tablehead{
		\colhead{Subsample}                      &
		\colhead{$\#$ galaxies}                   &
		\colhead{Cut 1}             &
		\colhead{Cut 2}                    &
		\colhead{$\braket{\logtenmv}$} &
		\colhead{$\langle z_{\rm gal} \rangle$} &
		\colhead{$\Delta v(50)$}                      &
		\colhead{$\Delta v(90)$}
	}

	\startdata

        \multicolumn{2}{c}{} &
	\multicolumn{3}{c}{Figure~\ref{fig:unnorm_masscutovi}: Absorber Kinematics} &
        \multicolumn{1}{c}{} &
        \multicolumn{2}{c}{$v=v_{\rm pix}$~({\kms})} \\
	\noalign{\vskip 1.0mm}
	\tableline \\[-4pt]
	Lower Mass  & 10 & ${\logtenmv} < 11.7$    & $\ldots$ & $11.62\pm0.09$ & $0.21\pm0.03$ &  $80_{-8}^{+5}$ & $188_{-19}^{+11}$ \\
	Higher Mass & 21 & ${\logtenmv} \geq 11.7$ & $\ldots$ & $12.00\pm0.04$ & $0.32\pm0.03$ & $108_{-9}^{+6}$ & $255_{-22}^{+14}$ \\
	Group~\tablenotemark{a} & 6 & $\ldots$                    & $\ldots$ & $\ldots$       & $0.19\pm0.05$ &  $64_{-7}^{+9}$ & $153_{-18}^{+21}$ \\[5pt]

        \tableline \\[-5pt]
        \multicolumn{2}{c}{} &
	\multicolumn{3}{c}{Figure~\ref{fig:masscutovi}: Absorber--Galaxy Kinematics} &
        \multicolumn{1}{c}{} &
        \multicolumn{2}{c}{$v=v_{\rm pix-gal}/V_{\rm c}(D)$} \\
	\noalign{\vskip 1.0mm}
	\tableline \\[-4pt]
	Lower Mass  & 15 & ${\logtenmv} < 11.88$    & $\ldots$ & $11.64\pm0.04$ & $0.22\pm0.03$ & $0.45_{-0.03}^{+0.03}$ & $1.12_{-0.09}^{+0.08}$ \\
	Higher Mass & 16 & ${\logtenmv} \geq 11.88$ & $\ldots$ & $12.06\pm0.07$ & $0.34\pm0.05$ & $0.34_{-0.02}^{+0.02}$ & $0.87_{-0.07}^{+0.06}$ \\[5pt]

        \tableline \\[-5pt]
        \multicolumn{2}{c}{} &
	\multicolumn{3}{c}{Figure~\ref{fig:zgalcutovi}: Absorber--Galaxy Kinematics} &
        \multicolumn{1}{c}{} &
        \multicolumn{2}{c}{$v=v_{\rm pix-gal}/V_{\rm c}(D)$} \\
	\noalign{\vskip 1.0mm}
	\tableline \\[-4pt]
	Lower-z & 15 & $z_{\rm{gal}} < 0.244$ & $\ldots$ & $11.73\pm0.18$ & $0.205\pm0.016$ & $0.40_{-0.04}^{+0.04}$ & $1.11_{-0.12}^{+0.11}$ \\
	Higher-z & 16 & $z_{\rm{gal}} \geq 0.244$ & $\ldots$ & $11.95\pm0.05$ & $0.39\pm0.03$ & $0.37_{-0.02}^{+0.02}$ & $0.92_{-0.05}^{+0.04}$ \\[5pt]

        \tableline \\[-5pt]
        \multicolumn{2}{c}{} &
	\multicolumn{3}{c}{Not Plotted: Absorber--Galaxy Kinematics} &
        \multicolumn{1}{c}{} &
        \multicolumn{2}{c}{$v=v_{\rm pix-gal}/V_{\rm c}(D)$} \\
	\noalign{\vskip 1.0mm}
	\tableline \\[-4pt]
	Face-on & $15$ &  $i<51^\circ$ & $\ldots$ & $11.82\pm0.14$ & $0.32\pm0.09$ & $0.40_{-0.04}^{+0.03}$ & $1.03_{-0.12}^{+0.09}$ \\
	Edge-on & $16$ & $i\geq51^\circ$ & $\ldots$ & $11.91\pm0.08$ & $0.24\pm0.05$ & $0.38_{-0.03}^{+0.02}$ & $0.96_{-0.08}^{+0.07}$ \\[5pt]
	\noalign{\vskip 1.0mm}
	\tableline \\
%                   &      &                &          &                &               &                        &                      \\
	Major Axis & $17$ & $\Phi<45^\circ$ & $\ldots$ & $11.76\pm0.16$ & $0.23\pm0.06$ & $0.35_{-0.03}^{+0.02}$ & $0.90_{-0.09}^{+0.06}$ \\
	Minor Axis & $14$ & $\Phi\geq45^\circ$ & $\ldots$ & $11.92\pm0.06$ & $0.33\pm0.05$ & $0.42_{-0.03}^{+0.04}$ & $1.07_{-0.10}^{+0.10}$ \\[5pt]

        \tableline \\[-5pt]
        \multicolumn{2}{c}{} &
	\multicolumn{3}{c}{Figure~\ref{fig:azimuthorienovi}: Absorber--Galaxy Kinematics} &
        \multicolumn{1}{c}{} &
        \multicolumn{2}{c}{$v=v_{\rm pix-gal}/V_{\rm c}(D)$} \\
	\noalign{\vskip 1.0mm}
	\tableline \\[-4pt]        
	Major Axis + Face-on & $10$ & $\Phi < 45^\circ$ & $i < 51^\circ$ & $12.00\pm0.12$ & $0.32\pm0.05$ & $0.35_{-0.03}^{+0.02}$ & $0.87_{-0.09}^{+0.06}$ \\
	Major Axis + Edge-on & $7$ & $\Phi < 45^\circ$ & $i \geq 51^\circ$ & $11.69\pm0.32$ & $0.26\pm0.07$ & $0.38_{-0.05}^{+0.06}$ & $0.97_{-0.17}^{+0.20}$ \\
	Minor Axis + Face-on & $5$ & $\Phi \geq 45^\circ$ & $i < 51^\circ$ & $11.64\pm0.17$ & $0.18\pm0.05$ & $0.51_{-0.06}^{+0.04}$ & $1.24_{-0.16}^{+0.10}$ \\
	Minor Axis + Edge-on & $9$ & $\Phi \geq 45^\circ$ & $i \geq 51^\circ$ & $11.96\pm0.13$ & $0.32\pm0.08$ & $0.38_{-0.03}^{+0.02}$ & $0.96_{-0.08}^{+0.07}$ \\[5pt]

        \tableline \\[-5pt]
        \multicolumn{2}{c}{} &
	\multicolumn{3}{c}{Figure~\ref{fig:mvazimuthorienovi}: Absorber--Galaxy Kinematics} &
        \multicolumn{1}{c}{} &
        \multicolumn{2}{c}{$v=v_{\rm pix-gal}/V_{\rm c}(D)$} \\
	\noalign{\vskip 1.0mm}
	\tableline \\[-4pt]        
	Major Axis + Lower Mass & $7$ & $\Phi < 45^\circ$ & ${\logtenmv} < 11.88$ &  $11.66\pm0.15$ & $0.31\pm0.09$ & $0.40_{-0.04}^{+0.04}$ & $1.02_{-0.16}^{+0.15}$ \\
	Major Axis + Higher Mass & $10$ & $\Phi < 45^\circ$ &${\logtenmv} \geq 11.88$ & $12.09\pm0.09$ & $0.29\pm0.06$ & $0.33_{-0.03}^{+0.02}$ & $0.83_{-0.09}^{+0.06}$ \\
	Minor Axis + Lower Mass & $8$ & $\Phi \geq 45^\circ$ & ${\logtenmv} < 11.88$ & $11.64\pm0.05$ & $0.20\pm0.02$ & $0.48_{-0.04}^{+0.03}$ & $1.17_{-0.11}^{+0.07}$ \\
	Minor Axis + Higher Mass & $6$ & $\Phi \geq 45^\circ$ & ${\logtenmv} \geq 11.88$ & $12.04\pm0.14$ & $0.41\pm0.07$ & $0.37_{-0.03}^{+0.03}$ & $0.92_{-0.11}^{+0.10}$ \\
	\noalign{\vskip 2.0mm}
	\tableline \\
	Face-on + Lower Mass & $6$ & $i < 51^\circ$ & ${\logtenmv} < 11.88$ & $11.65\pm0.09$ & $0.22\pm0.08$ & $0.51_{-0.05}^{+0.03}$ & $1.23_{-0.14}^{+0.08}$ \\
	Face-on + Higher Mass & $9$ & $i < 51^\circ$ & ${\logtenmv} \geq 11.88$ & $12.00\pm0.15$ & $0.30\pm0.06$ & $0.33_{-0.02}^{+0.03}$ & $0.81_{-0.06}^{+0.08}$ \\
	Edge-on + Lower Mass & $9$ & $i \geq 51^\circ$ & ${\logtenmv} < 11.88$ & $11.64\pm0.10$ & $0.23\pm0.04$ & $0.40_{-0.03}^{+0.03}$ & $0.99_{-0.10}^{+0.11}$ \\
	Edge-on + Higher Mass & $7$ & $i \geq 51^\circ$ & ${\logtenmv} \geq 11.88$ & $12.13\pm0.08$ & $0.39\pm0.08$ & $0.37_{-0.03}^{+0.03}$ & $0.93_{-0.10}^{+0.09}$ \\[-5pt]
	\enddata
        \tablenotetext{a}{Data from \citet{pointon17}.}
\end{deluxetable*}

\begin{figure}[t]
	\includegraphics[width=\linewidth]{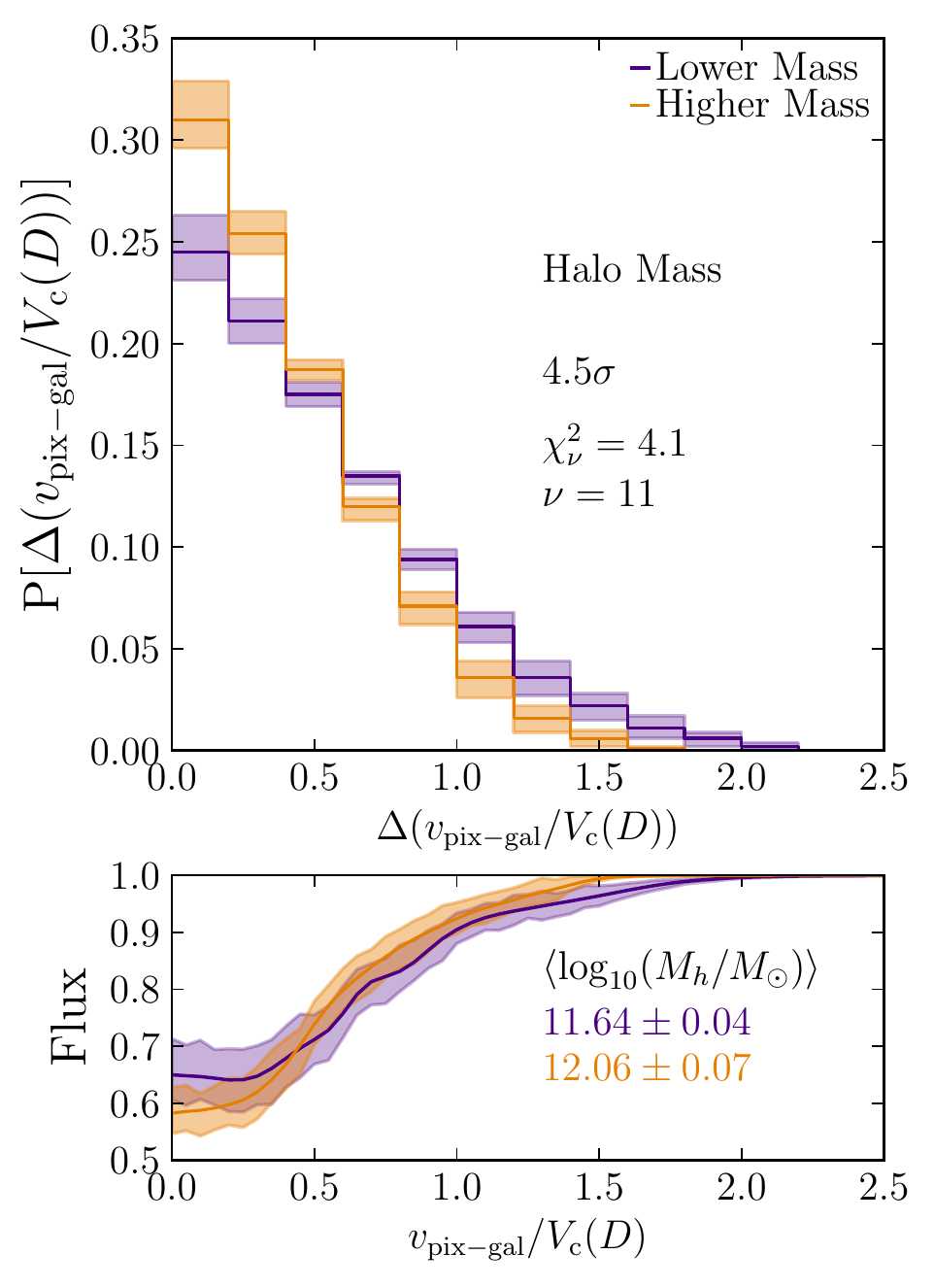}
        \caption[]{(Top) Normalized absorber--galaxy TPCFs for lower
          mass and higher mass galaxies. Unlike
          Figure~\ref{fig:unnorm_masscutovi}, the pixel--galaxy
          velocities are normalized by the circular velocity at the
          observed impact parameter, $V_{\rm c}(D)$, to account for
          the range of halo masses in the sample. This shows a strong
          mass-dependence for the kinematics associated with {\OVI} at
          the $4.5\sigma$ level, where lower mass galaxies have
          significantly larger velocity dispersions than higher mass
          galaxies. (Bottom) The average absorption spectra
          corresponding to lower mass galaxies and higher mass
          galaxies.~$1\sigma$ uncertainties for both the TPCFs and
          average absorption spectra were calculated with a bootstrap
          analysis with 100 bootstrap realizations.}
	\label{fig:masscutovi}
\end{figure}

\begin{figure}[t]
	\includegraphics[width=\linewidth]{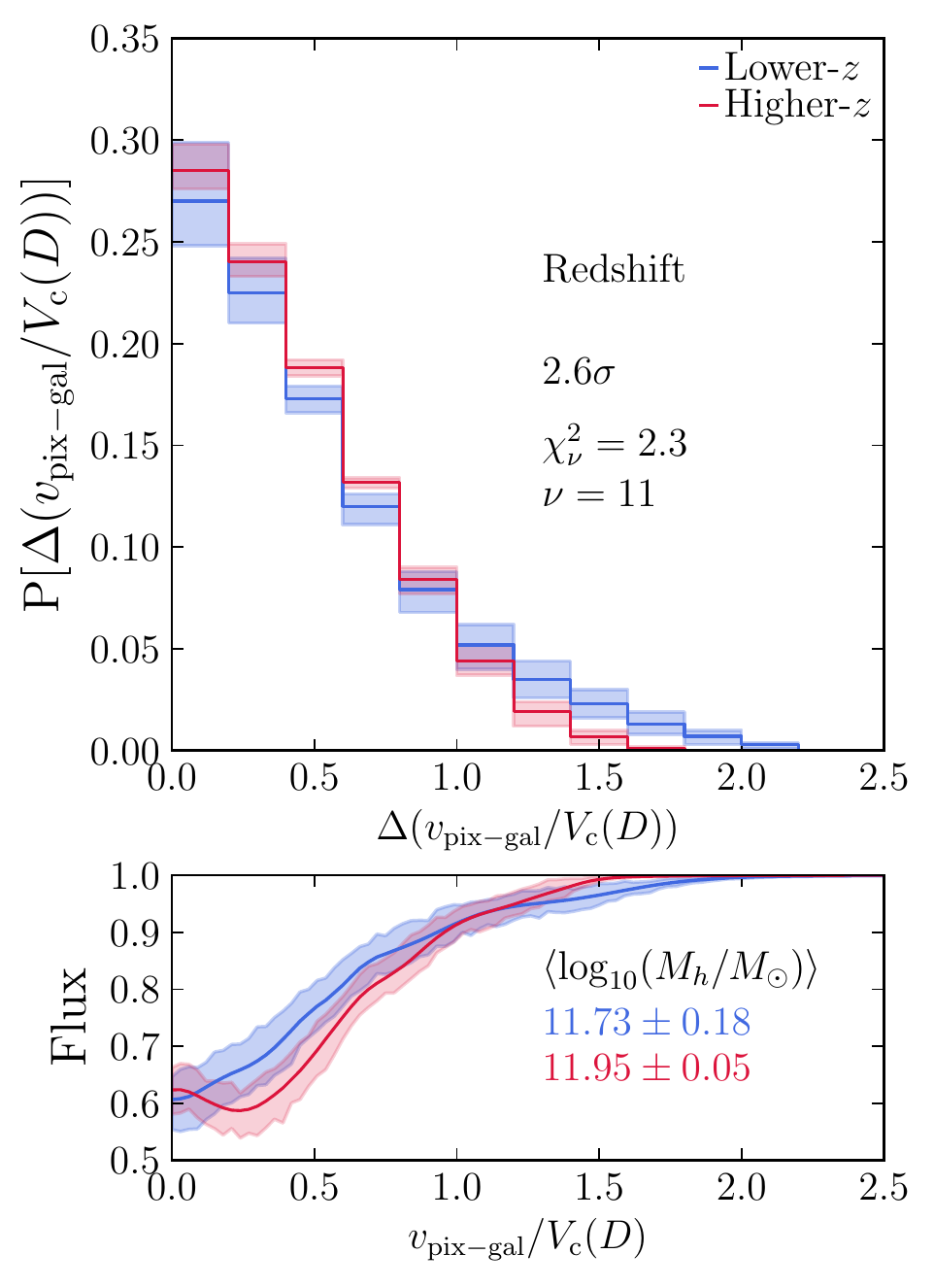}
	\caption[]{(Top) Normalized absorber--galaxy TPCFs for lower
          redshift and higher redshift galaxies. There is a probable
          redshift dependence of the {\OVI} kinematics at the
          $2.6\sigma$ level, where {\OVI} absorbing gas in lower
          redshift galaxies has larger velocities relative to {\OVI}
          absorbing gas in higher redshift galaxies. (Bottom) The
          average absorption spectra corresponding to lower and higher
          redshift galaxies. $1\sigma$ uncertainties for both the
          TPCFs and the average spectra were calculated with a
          bootstrap analysis with 100 bootstrap realizations.}
	\label{fig:zgalcutovi}
\end{figure}

\begin{figure*}[t]
	\includegraphics[width=\textwidth]{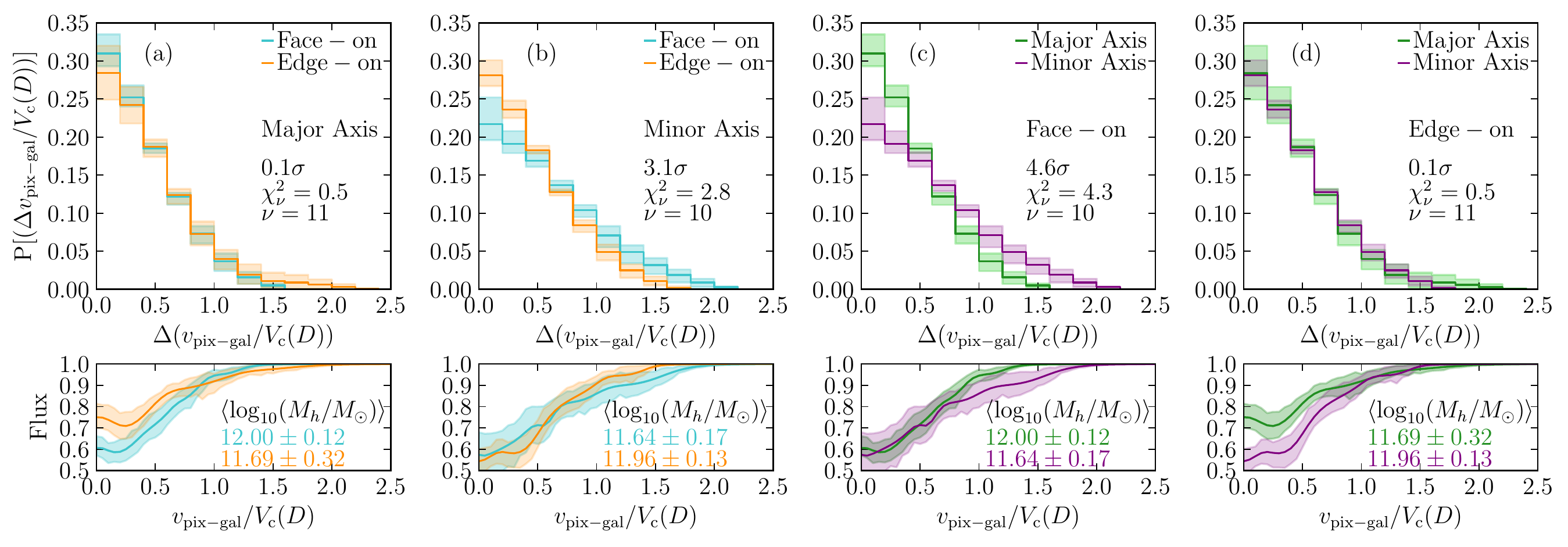}
	\caption[]{(Top) Normalized absorber--galaxy TPCFs for a)
          edge-on and face-on galaxies probed along the major axis; b)
          edge-on and face-on galaxies probed along the minor axis; c)
          face-on galaxies probed along the major and minor axes; d)
          edge-on galaxies probed along the major and minor
          axes. (Bottom) Average absorption profiles corresponding to
          the top panels. In a) and b), cyan corresponds to face-on
          galaxies, and orange corresponds to edge-on galaxies. In c)
          and d), green corresponds to galaxies probed along the major
          axis, and purple corresponds to galaxies probed along the
          minor axis. The uncertainties for the average absorption
          profiles were calculated with a bootstrap analysis with 100
          bootstrap realizations.}
	\label{fig:azimuthorienovi}
\end{figure*}

\section{Absorber--Galaxy Kinematics}
\label{sec:abs-galkin}

In this section we examine the relative absorber--galaxy kinematics
using the TPCFs described in Section~\ref{sec:TPCFabs-gal}. Note that
we normalize pixel--galaxy velocities by the circular velocity at the
observed impact parameter, $V_{\rm c}(D)$, in order to assess the
kinematics in a mass-independent way. This is done in light of the
results of the previous section, which showed a significant
mass-dependence in the absorber kinematics. We also only focus on
isolated galaxy sightlines (i.e., no group environment kinematics from
Figure~\ref{fig:unnorm_masscutovi}) due to the velocity
normalization. Since the {\OVI} gas is in an intra-group medium and is
not specifically associated with a particular galaxy, we cannot assign
a single $V_{\rm c}(D)$ for each group absorber.
$ $ \\

\subsection{Bivariate Analysis}
\label{sec:bivariate-absgal}

We again slice the sample by mass, but choose the median halo mass of
the sample ($\braket{\logtenmv}=11.88$) in order to have roughly equal
numbers of galaxies in each mass bin for a later multivariate
analysis. The absorber--galaxy TPCFs and associated average absorption
profiles for lower mass and higher mass galaxies are plotted in
Figure~\ref{fig:masscutovi}. After accounting for the inherent mass
bias in the absorber kinematics above, we find that lower mass
galaxies ($\logtenmv< 11.88$) have larger absorber--galaxy velocity
dispersions compared with higher mass galaxies ($\logtenmv\geq 11.88$)
at a significance of $4.5\sigma$. From Table~\ref{tab:tpcfprops_OVI},
$\Delta v(50)$ and $\Delta v(90)$ for lower mass galaxies
($0.45_{-0.03}^{+0.03}$ and $1.12_{-0.09}^{+0.08}$, respectively) are
larger than that for higher mass galaxies ($0.34_{-0.02}^{+0.02}$ and
$0.87_{-0.07}^{+0.06}$, respectively).

The average absorption profiles below the TPCFs in
Figure~\ref{fig:masscutovi} show that the bulk of the absorption lies
around the galaxy systemic velocity, and this result applies to all
subsequent subsample slices. Furthermore, most of the absorption lies
within the host galaxy's circular velocity at the observed impact
parameter, indicating that this gas is likely bound to the galaxy. In
a complement to the TPCFs, the average absorption profile for lower
mass galaxies extends to larger normalized pixel--galaxy velocities
than for higher mass galaxies. This small fraction of the absorption
profiles has velocities greater than $V_c(D)$, indicating that this
gas may not be bound to the host galaxies, and the fraction may be
larger for lower mass galaxies.

To account for any redshift evolution of the absorber--galaxy
kinematics, we cut the sample by the median redshift, $\braket{z_{\rm
    gal}} = 0.244$. This was motivated by work done in
\citet{magiicat4}, who found that the {\MgII} absorber velocity
dispersions for red galaxies decreased with decreasing redshift,
possibly indicative of the quenching of star formation. In
Figure~\ref{fig:zgalcutovi}, there is a weak suggestion that lower
redshift galaxies ($\braket{z_{\rm gal}} < 0.244$) have larger
absorber--galaxy velocity dispersions than higher redshift galaxies
($\braket{z_{\rm gal}} \geq 0.244$) at a significance of
$2.6\sigma$. This result trends in the opposite direction as that
found with {\MgII}, although here we are investigating the relative
absorber--galaxy velocity dispersions rather than absorber velocity
dispersions. Referring to Table~\ref{tab:tpcfprops_OVI}, $\Delta
v(50)$ measurements for lower redshift galaxies and higher redshift
galaxies are similar, although $\Delta v(90)$ for lower redshift
galaxies is slightly larger than for higher redshift galaxies. The
median masses for the two subsamples are consistent within
uncertainties, so any differences seen here are likely not dominated
by the mass dependence of Figure~\ref{fig:masscutovi}. The average
absorption profiles in the bottom panel of Figure~\ref{fig:zgalcutovi}
show that lower redshift galaxies have a nontrivial optical depth at a
larger normalized pixel--galaxy velocity compared with the higher
redshift galaxies, but the two profiles are still consistent within
uncertainties. It is possible that {\OVI} kinematics have a redshift
dependence, analogous to that of {\MgII} kinematics in red galaxies
\citep{magiicat4}, but this investigation would be better done on a
sample with a greater range in redshifts (up to $z\sim2-3$).

We also cut the sample by the median inclination, $\braket{i} =
51^\circ$, where galaxies with $i\geq51^\circ$ are considered
``edge-on" galaxies and galaxies with $i<51^\circ$ are considered
``face-on" galaxies. We find that both edge-on and face-on galaxies
have the same velocity dispersions ($0.01\sigma$, not plotted) and
average absorption profiles. The $\Delta v(50)$ and $\Delta v(90)$
measurements reported in Table~\ref{tab:tpcfprops_OVI} are consistent
within uncertainties. The median masses and median redshifts for the
two subsamples are consistent within uncertainties.

A final bivariate cut made to the sample is the median azimuthal angle
of $\braket{\Phi} = 45^\circ$, where galaxies with $\Phi\geq45^\circ$
are considered ``minor axis" galaxies, that is, galaxies that are
probed along their projected minor axis. ``Major axis" galaxies, where
$\Phi<45^\circ$, are defined in a similar fashion. We find that minor
axis galaxies have similar velocity dispersions to major axis galaxies
($1.5\sigma$, not plotted), and the average absorption profiles are
not qualitatively different. From Table~\ref{tab:tpcfprops_OVI}, the
$\Delta v(50)$ and $\Delta v(90)$ measurements for both major axis
galaxies and minor axis galaxies are inconsistent within
uncertainties, but are still within $1.4\sigma$. Additionally, the
median masses and median redshifts for the major axis galaxies and
minor axis galaxies are consistent within uncertainties.

The independence of the {\OVI} kinematics on the inclination and
azimuthal angles reported here has been confirmed in previous work
\citep{nielsen17, kacprzak19}, but here we have obtained the same
conclusion with absorber--galaxy kinematics for simple bivariate cuts.

\subsection{Multivariate Analysis} \label{sec:multivariate}

\subsubsection{Inclination and Azimuthal Angle} \label{sec:321}

In Figure~\ref{fig:azimuthorienovi}, we present the TPCFs for galaxy
subsamples cut by inclination, $i$, as well as galaxy azimuthal angle,
$\Phi$, for all halo masses. The corresponding average absorption
profiles are plotted in the panels below the TPCFs.

In Figure~\ref{fig:azimuthorienovi}(a), we compare face-on (cyan
lines) and edge-on (orange lines) galaxies probed along their
projected major axis. There is no difference in the absorber--galaxy
velocity dispersions between the two inclinations ($0.1\sigma$). The
associated $\Delta v(50)$ and $\Delta v(90)$ measurements presented in
Table~\ref{tab:tpcfprops_OVI} are consistent within uncertainties,
providing further support that the velocity dispersions between
face-on and edge-on inclinations for major axis galaxies are very
similar. The median masses for the two subsamples are consistent
within uncertainties. The average absorption profile shows that there
is a slight tail for the edge-on major axis galaxies at higher
normalized pixel--galaxy velocities. While this is reflected in the
TPCFs, the difference between the plotted subsamples is not
significant due to large uncertainties. There is also more optical
depth near the galaxy systemic velocity for the face-on major axis
galaxies compared with the edge-on major axis galaxies.

Figure~\ref{fig:azimuthorienovi}(b) compares face-on (cyan lines) and
edge-on (orange lines) inclinations as well, but for galaxies probed
along their projected minor axis. There is a significant difference in
the velocity dispersions between the two inclinations ($3.1\sigma$),
where face-on galaxies probed along the minor axis have a slightly
larger velocity dispersion than edge-on galaxies. From
Table~\ref{tab:tpcfprops_OVI}, the associated $\Delta v(50)$
measurements for face-on minor axis galaxies ($0.51_{-0.06}^{+0.04}$)
are larger than for edge-on minor axis galaxies
($0.38_{-0.03}^{+0.02}$). Additionally, $\Delta v(90)$ for face-on
minor axis galaxies ($1.24_{-0.16}^{+0.10}$) is larger than for
edge-on minor axis galaxies ($0.96_{-0.08}^{+0.07}$). This may be
attributable to the observation that edge-on minor axis galaxies
($\braket{\logtenmv} = 11.96\pm0.13$) tend to be of higher mass than
that of face-on minor axis galaxies ($\braket{\logtenmv} =
11.64\pm0.17$), so the corresponding halos have a larger virial
temperature. This provides the conditions for {\OVI} to ionize into
higher order species and thus reduces the size and kinematic extent of
the {\OVI} clouds \citep{oppenheimer16, pointon17}. The average
absorption profiles clearly show that there is a larger optical depth
at higher normalized velocities for the face-on minor axis galaxies
compared with the edge-on minor axis galaxies. This is seen as the
tail in the TPCF, but the uncertainties in the TPCFs are large. The
average absorption profile shows no differences in the absorption
between the two subsamples around the galaxy systemic velocity.

Figure~\ref{fig:azimuthorienovi}(c) compares face-on galaxies probed
along the major axis (green lines) and minor axis (purple
lines). There is a very significant difference ($4.6\sigma$) in the
velocity dispersions, with minor axis galaxies showing a larger
velocity dispersion compared with major axis galaxies in face-on
($i<51^\circ$) inclinations. This is also seen in the $\Delta v(50)$
and $\Delta v(90)$ measurements reported in
Table~\ref{tab:tpcfprops_OVI}, where $\Delta v(50)$ for face-on
galaxies probed along the minor axis ($0.51_{-0.06}^{+0.04}$) is
significantly larger than when probed along the major axis
($0.35_{-0.03}^{+0.02}$). Likewise, the $\Delta v(90)$ value for
face-on galaxies probed along the minor axis ($1.24_{-0.16}^{+0.10}$)
is significantly larger than face-on galaxies probed along the major
axis ($0.87_{-0.09}^{+0.06}$). The median mass for face-on minor axis
galaxies, $\braket{\logtenmv} = 11.64\pm0.17$, is lower than that for
face-on major axis galaxies, $\braket{\logtenmv} = 12.00\pm0.12$,
where the direction of the difference reflects the result of
Figure~\ref{fig:masscutovi}. The mass difference likely plays a role
in the significant difference in the TPCFs, rather than being only an
inclination effect. The average absorption profile also shows a much
larger optical depth at higher normalized velocities (largely for
$v_{\rm pix-gal}/V_{\rm c}(D) > 1$, where the gas is more likely to be
unbound) for face-on minor axis galaxies compared with the face-on
major axis galaxies, which is seen as the tail in the TPCF. Where the
gas is expected to be bound, there is no difference in the average
absorption profiles around the galaxy systemic velocity.

\begin{figure*}[htp]
	\includegraphics[width=\textwidth]{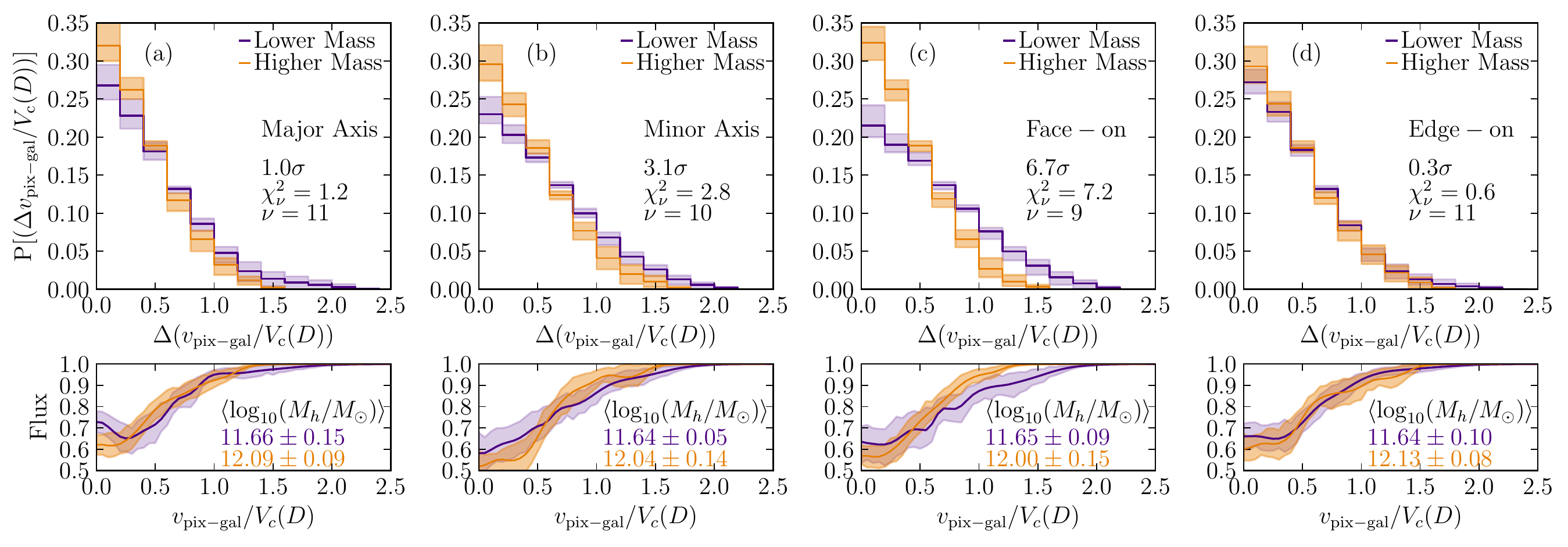}
	\caption[]{(Top) Normalized absorber--galaxy TPCFs for lower
          mass and higher mass galaxies probed along a) the major
          axis; b) the minor axis, and seen in c) face-on
          inclinations; d) edge-on inclinations. (Bottom) Average
          absorption profiles corresponding to the top panels. Orange
          corresponds to higher mass galaxies, and purple corresponds
          to lower mass galaxies. The uncertainties were calculated
          with a bootstrap analysis with 100 bootstrap realizations.}
	\label{fig:mvazimuthorienovi}
\end{figure*}

Finally, Figure~\ref{fig:azimuthorienovi}(d) compares edge-on galaxies
probed along the major axis (green lines) and minor axis (purple
lines). There is no difference ($0.1\sigma$) in the velocity
dispersions between the two azimuthal angle categories. This is
reflected in the overlap of the respective $\Delta v(50)$ and $\Delta
v(90)$ measurements reported in Table~\ref{tab:tpcfprops_OVI}. The
median masses for edge-on major axis galaxies and edge-on minor axis
galaxies are consistent within uncertainties. The average absorption
profiles also have similar optical depths at the larger normalized
velocities, but there is a larger optical depth for the edge-on minor
axis galaxies around the galaxy systemic velocity compared with the
edge-on major axis galaxies.

\subsubsection{Halo Masses and Orientation}
\label{sec:322}

Given the significant difference in the absorber--galaxy kinematics
for lower mass galaxies compared to higher mass galaxies, the results
in the previous section with galaxy inclinations and azimuthal angles
may largely be due to the distribution of galaxy masses in each
subsample comparison. In Figure~\ref{fig:mvazimuthorienovi}, we
present TPCFs for galaxy subsamples cut by halo mass, {\logtenmv}, and
one of two galaxy orientation measures: inclination, $i$, or azimuthal
angle, $\Phi$. The corresponding average absorption profiles are
plotted in the panels below the TPCFs. We conduct this test to better
pinpoint the galaxy properties that are most important in governing
the absorber--galaxy kinematics.

Figure~\ref{fig:mvazimuthorienovi}(a) compares lower mass (indigo
lines) and higher mass (orange lines) galaxies probed along the major
axis. There is no difference ($1.0\sigma$) in the velocity
dispersions, though the lower mass galaxies have a slightly wider
velocity dispersion tail (but with large uncertainties) than higher
mass galaxies when probed along the major axis. This is reflected in
the consistency, within uncertainties, of the $\Delta v(50)$ and
$\Delta v(90)$ measurements from Table~\ref{tab:tpcfprops_OVI} between
the two subsamples. The average absorption profiles are also
consistent within uncertainties.

Figure~\ref{fig:mvazimuthorienovi}(b) shows the TPCFs for minor axis
galaxies. There is a significant difference ($3.1\sigma$) between the
velocity dispersions of the lower mass (indigo lines) and higher mass
(orange lines) galaxies, where lower mass galaxies have a larger
velocity dispersion than higher mass galaxies probed along the minor
axis. The difference in velocity dispersions is also seen in the
$\Delta v(50)$ and $\Delta v(90)$ measurements reported in
Table~\ref{tab:tpcfprops_OVI}, where these values for lower~mass
galaxies ($0.48_{-0.04}^{+0.03}$ and $1.17_{-0.11}^{+0.07}$,
respectively) are larger than for higher mass galaxies
($0.37_{-0.03}^{+0.03}$ and $0.92_{-0.11}^{+0.10}$,
respectively). While this difference was expected from
Figure~\ref{fig:masscutovi}, the fact that the comparison for galaxies
probed along the projected major axis in
Figure~\ref{fig:mvazimuthorienovi}(a) is insignificant indicates that
azimuthal angle plays a role in the observed kinematic
structure. However, the average absorption profiles for the two
subsamples are comparable within uncertainties.

In Figure~\ref{fig:mvazimuthorienovi}(c), we compare lower and higher
mass TPCFs for face-on galaxies. There is a very significant
difference ($6.7\sigma$) between the velocity dispersions of the lower
mass (indigo lines) and higher mass (orange lines) galaxies, where
lower mass face-on galaxies have a much larger velocity dispersion
than higher mass face-on galaxies. The difference in velocity
dispersions is reflected in the $\Delta v(50)$ and $\Delta v(90)$
measurements from Table~\ref{tab:tpcfprops_OVI}, where the values for
the lower mass face-on galaxies ($0.51_{-0.05}^{+0.03}$ and
$1.23_{-0.14}^{+0.08}$, respectively) are much greater than for higher
mass, face-on galaxies ($0.33_{-0.02}^{+0.03}$ and
$0.81_{-0.06}^{+0.08}$). The average absorption profiles also show
that the optical depth at higher normalized velocities is much larger
for the lower mass face-on galaxies compared with the higher mass
face-on galaxies.

In Figure \ref{fig:mvazimuthorienovi}(d), we compare the TPCFs of
lower mass and higher mass galaxies with edge-on inclinations.~There
is no difference ($0.3\sigma$) between the velocity dispersions of the
lower mass edge-on (indigo lines) and higher mass edge-on (orange
lines) galaxies. This is seen in the consistency between their $\Delta
v(50)$ and $\Delta v(90)$ measurements. The mass-dependence found in
Figure~\ref{fig:masscutovi} is not present for edge-on galaxies, but
is for face-on galaxies in the previous paragraph, suggesting that
galaxy inclination is important for determining the absorber--galaxy
kinematics. The average absorption profiles of the lower mass edge-on
galaxies and the higher mass edge-on galaxies are very similar, where
the bulk of the absorption occurs around the galaxy systemic velocity,
and the optical depth tapers off at similar normalized velocities.

Finally, we tested similar mass bins for the orientation subsamples,
although these comparisons are not plotted directly. There are no
significant differences between major and minor axis galaxies for
either the lower mass ($0.8\sigma$) or higher mass ($0.4\sigma$)
subsamples. Similarly, there is no significant difference between
edge-on and face-on galaxies for the higher mass subsample. However,
lower mass face-on galaxies have significantly larger velocity
dispersions compared to edge-on lower mass galaxies ($3.3\sigma$). In
this section, the clear outlier in kinematics is thus the lower mass
face-on subsample.

\section{Discussion}
\label{sec:discussion}

From the TPCFs and average absorption profiles presented above, we
find that there is a strong mass dependence of the {\OVI} absorber
kinematics, where higher mass (${\logtenmv} \geq 11.7$) isolated
galaxies have larger velocity dispersions compared with lower mass
(${\logtenmv} < 11.7$) isolated galaxies. Group galaxies have much
narrower velocity dispersions than the higher mass isolated galaxies
($13.1\sigma$). We also find that the absorber--galaxy kinematics
display non-virialized motions. These are primarily due to outflows in
face-on and minor axis orientations for lower mass galaxies. These
motions were found after normalizing the pixel--galaxy velocities by
the circular velocity at the observed impact parameter, $V_{\rm
  c}(D)$, to account for the range of halo masses in the sample.

\subsection{Absorber Kinematics}
\label{sec:absdiscussion}

The mass dependence of the TPCFs and average absorption profiles in
Figure \ref{fig:unnorm_masscutovi} may be attributable to the strength
of the {\OVI} absorption, quantified by the column density, $\log
N_{\tiny \OVI}$.~Simulations of {\OVI} in galaxy halos show that the
column density reaches its apex at ${\logtenmv}=12$ because the virial
temperature of these galaxies is comparable to the temperature at
which the {\OVI} ionization fraction is greatest
(\citealp{oppenheimer16}; also see \citealp{nelson18} and
\citealp{oppenheimer18} for an alternative explanation). Galaxy halos
with lower (${\logtenmv}<11.7$) or higher masses (${\logtenmv}>12.3$,
i.e., group environments) have lower column densities or weaker {\OVI}
absorption. This is likely because a smaller (larger) galaxy halo mass
corresponds to a halo that is too cold (hot) for a significant
fraction of {\OVI}. We confirm this trend in the observational column
densities plotted in Figure~\ref{fig:colopp}. This relation may also
lead to smaller {\OVI} clouds and a statistically lower kinematic
extent for sub-$L^{\ast}$ and group galaxies.

We clearly see this effect in the absorber kinematics. The median mass
for higher mass galaxies corresponds to $\sim L^*$ galaxies, which
have a virial temperature that is most conducive for the presence of
{\OVI} gas according to \citet{oppenheimer16}. Our results indicate
that this also leads to a larger kinematic extent of the {\OVI} gas
since there is more gas to be distributed along the line of sight. In
contrast, lower mass galaxies at sub-$L^*$ likely have virial
temperatures which are too cool for a significant presence of {\OVI},
resulting in more narrow kinematic dispersions.~As shown by
\citet{pointon17} and confirmed here, group galaxies are the other
extreme in that they likely live in halos with virial temperatures
that are too warm for {\OVI}.~As {\OVI} ionizes out into higher order
species (e.g., \OVII\ and beyond) for these highest mass galaxies,
this translates to narrower kinematic extents because there is less
gas along the line of sight at the temperatures required for strong
{\OVI} absorption.

We also saw this mass--{\OVI} effect in the optical depth distribution
of the average absorption profiles, constructed from the {\OVI}
$\lambda1031$ transition. For all pixel velocities, higher mass
galaxies (${\logtenmv}\approx12$) have more absorption compared to
both lower mass galaxies ($\logtenmv < 11.7$) and group environments
($\logtenmv\gtrsim12.3$). As shown in Figure~\ref{fig:colopp}, the
column densities obtained from these average absorption profiles,
$\log N_{\tiny \OVI}=14.44$, 14.57, and 14.18 for increasing halo
mass, follow a similar trend with mass as the aperture column
densities obtained by \citet{oppenheimer16} in the EAGLE
simulations.\footnote{There is a well-known tension between
  observations and simulations, where simulations under-produce {\OVI}
  absorption \citep[e.g.,][]{oppenheimer16, suresh17}, and this is the
  cause of the offset between the observed and simulated points.}

Previous absorber kinematics work by \citet{nielsen17} did not
investigate this mass dependence, rather they focused on the star
formation properties of the galaxies with $B-K$ colors. They found
that the kinematics did not depend on galaxy orientations or star
formation activity and suggested that the observed gas may be a result
of ancient outflows which have had time to form a roughly
kinematically uniform halo. They also suggested that differing
ionization conditions throughout the CGM result in the azimuthal angle
dependence of {\OVI} found by \citet{kacprzak15}. Combined with the
results presented here, we suggest that the absorption properties of
{\OVI} are not good straightforward probes of current baryon cycle
processes in galaxies. Instead, {\OVI} absorption properties are
primarily probes of halo mass.

\subsection{Absorber--Galaxy Kinematics}

In an attempt to better pull out baryon cycle signatures in the {\OVI}
absorption profiles, we shifted the absorption relative to the galaxy
systemic velocity and normalized these pixel--galaxy velocities by the
galaxy circular velocity at the observed impact parameter,
$V_c(D)$. The resulting absorber--galaxy kinematics are effectively
independent of the trends found with the absorber kinematics. Thus the
differences between subsamples for the absorber--galaxy kinematics
should not be dominated by mass and, due to studying the relative
velocities between the observed gas and the host galaxies, may reflect
baryon cycle processes.

We first tested this by examining the relative absorber--galaxy
kinematics for lower and higher mass galaxies in
Figure~\ref{fig:masscutovi}. Overall, the bulk of the absorption lies
near the galaxy systemic velocity, with little gas exceeding the
galaxy circular velocity, similar to previous work comparing to the
galaxy escape velocity \citep{tumlinson11, mathes14, kacprzak19}. The
small fraction of gas that does exceed $V_c(D)$ is more likely to
escape the galaxy, and this fraction is slightly larger for lower mass
galaxies, which have statistically larger ($4.5\sigma$)
absorber--galaxy kinematics. \citet{mathes14} examined a sample of 11
galaxies with measured {\OVI} absorption and found that the gas around
lower mass galaxies is more likely to escape the halo than higher mass
galaxies. The authors suggested this is evidence for differential
kinematics due to differential wind recycling, where outflowing gas in
higher mass galaxies is more likely to recycle back onto the host
galaxy due to lower outflow velocities relative to $V_c(D)$
\citep{oppenheimer10}.

\citet{mathes14} also presented CGM radial velocities for a simulated
$\logtenmv \simeq11.3$ galaxy. Radially infalling gas rarely exceeded
$V_c(D)$ beyond 40~kpc due to the velocities being determined by the
galaxy's potential well, where the gas is largely subject to
gravitational forces inward and potentially (weaker) forces that slow
down the material as it travels toward the galaxy. Conversely,
outflowing gas often exceeded $V_c(D)$ out to at least 140~kpc. The
differences in the TPCF kinematics we present here are in the regime
where the gas flows exceed $V_c(D)$ at large impact parameters. Thus
we rule out accretion as the dominant source of this material and
further suggest that much of the {\OVI} traces outflowing gas.

We examined this further by studying the kinematics as a function of
galaxy orientation, where outflows are expected to dominate sightlines
probing galaxies along their projected minor axis or probing face-on
galaxies. Simple bivariate cuts with galaxy inclination or azimuthal
angle did not result in significant differences between the
kinematics. This is likely due to other factors such as comparable
halo masses between the subsamples where differential kinematics would
not be obvious. Additionally, outflows are not expected along the
projected major axis of edge-on galaxies, or for major axis galaxies
with inclinations closer to being face-on. Thus, we sliced the sample
further.

Accretion/rotation line-of-sight velocities are maximized for edge-on
galaxies along the major axis \citep[at least for low impact
  parameter, e.g.,][]{steidel02, ggk-sims, stewart11, danovich12,
  danovich15}, whereas outflow velocities are minimized
\citep[e.g.,][]{rubin-winds14}. Examining these orientations, we found
no significant difference in the absorber--galaxy TPCFs between
face-on galaxies and edge-on galaxies probed along their projected
major axes ($0.1\sigma$), as shown in Figure~\ref{fig:azimuthorienovi}(a). The TPCFs presented in Figure~\ref{fig:azimuthorienovi}(d), comparing major and minor axis
edge-on galaxies, are also consistent ($0.1\sigma$). The only
differences for these subsamples are in the average absorption
profiles near the galaxy systemic velocity, where larger optical
depths are found for the face-on galaxies or along the minor
axis. These latter differences are not due to a dependence on halo
mass since the subsamples in both comparisons have similar median halo
masses and the mass dependence is normalized out. Instead, these
results may be a signature of ancient outflows, which {\OVI} may trace
\citep[e.g.,][]{ford14}. In an edge-on inclination, outflows manifest
as minor axis absorption centered around the galaxy systemic velocity
since the radial velocity component is mostly perpendicular to a
galaxy's disk. We see this signature as more absorption along the
minor axis, and less along the major axis where accreting gas is
preferred.

A potential problem for interpreting {\OVI} gas origins is introduced
via the quasar sightline technique. Recently, \citet{kacprzak19}
examined the relative velocities between {\OVI} and the host galaxy
rotation curves for absorbers along the projected major axis. They
found that observed {\OVI} does not correlate with galaxy rotation and
has velocity dispersions that span the entire range of galaxy rotation
velocities. In contrast, the authors found inflowing {\OVI} filaments
in the CGM of simulated galaxies. The discrepancy between the observed
and simulated data is a result of the observational technique, where
the observed kinematic signatures of gas infall/rotation are blurred
due to multiple kinematic structures along the lines-of-sight which
cannot be disentangled with the data currently available \citep[also
  see][]{churchill15, peeples18}. Additionally, {\OVI} absorbing
clouds are predicted to be large, with radii on the order of tens to
hundreds of kiloparsecs from photoionization modeling \citep{lopez07,
  muzahid14, hussain15, stern16}.\footnote{Though note that the
  \citet{oppenheimer16} results assume {\OVI} is collisionally
  ionized. It is more likely that {\OVI} is some combination of
  photoionized and collisionally ionized.} Since accreting filaments
tend to have small covering fractions \citep[e.g.,][]{martin12,
  rubin-accretion}, gas infall signatures are overwhelmed by outflow
signatures.

We now focus on orientations in which outflows are expected to
dominate kinematic signatures: face-on galaxies and galaxies probed
along their projected minor axes. In
Figure~\ref{fig:azimuthorienovi}(b), there is a significant difference
between face-on and edge-on galaxies probed along the minor axis
($3.1\sigma$), where the velocity dispersions are larger for face-on
galaxies. The difference likely comes about because outflow velocities
are minimized in edge-on inclinations but are maximized for face-on
inclinations. The average absorption profiles are similar within
uncertainties, but there is a suggestion that the edge-on subsample is
more centrally concentrated near the galaxy systemic velocity, as
expected for a bipolar outflow geometry. The optical depth profile is
more smooth for the face-on subsample out to large normalized pixel
velocities, likely reflecting varying outflowing velocities with
impact parameter. This is supported by the results of
\citet{kacprzak19}, who found that minor axis {\OVI} gas can be
modeled by a decelerating outflow.

In another comparison, Figure~\ref{fig:azimuthorienovi}(c), we found
that face-on galaxies probed along the projected minor axis have
significantly larger velocity dispersions than face-on major axis
galaxies ($4.6\sigma$).\footnote{Recall that ``face-on'' here means
  $i<51^{\circ}$, which is quite inclined from a strict definition of
  ``face-on'', so the azimuthal angles are well-modeled in these
  systems.} One might expect outflows to dominate the signature in all
face-on galaxies, so this result seems unusual. However, the face-on
minor axis subsample has a larger impact parameter on average,
$\braket{D}=95.7$~kpc, than the face-on major axis subsample,
$\braket{D}=75.6$~kpc. We expect that outflows should influence the
observed gas more strongly at lower impact parameter
\citep[e.g.,][]{bordoloi11}, which is opposite the result we
find. Since the major axis subsample has significantly smaller
velocity dispersions at lower impact parameters, outflows do not
appear to dominate the face-on major axis absorption signature as
strongly as along the minor axis. Furthermore, the major axis
subsample has a larger median mass than the minor axis subsample. In
the differential kinematics/recycling scenario, less massive galaxies
can host larger $V_{\rm c}(D)$ normalized velocities due to smaller
potential wells, which is what we observe. The combination of large
absorber--galaxy velocity dispersions and large impact parameters
along the minor axis of face-on galaxies in
Figure~\ref{fig:azimuthorienovi}(c) is interesting given the
simulation results presented in \citet{kacprzak19}. The authors found
that for simulated galaxies at $z=1$ with $\logtenmv\sim11.7$, minor
axis {\OVI} outflows accelerate out to $D\sim 50$~kpc, where the gas
begins to decelerate and later falls back onto the host
galaxy. Perhaps the larger mean impact parameter for our face-on minor axis sample reflects the build-up of {\OVI} gas due to this
velocity turn-over, making the gas more easily observed.

The ease at which outflowing gas signatures are observed in lower mass
face-on galaxies over other subsamples in Figure~\ref{fig:mvazimuthorienovi}  could be explained with the
differential kinematics/wind recycling described by
\citet{oppenheimer10} and \citet{mathes14}. The specific star
formation rate is expected to be larger for lower mass galaxies, where
lower mass galaxies are more likely to be actively creating their
stellar populations from recent star formation
\citep[e.g.,][]{feulner05}. In this instance, outflowing gas in lower
mass galaxies is likely more recent, which reduces the amount of time
available for {\OVI} gas to kinematically mix within the CGM (evidence
for this kinematic mixing of {\OVI} is discussed in
\citealp{nielsen17} and ``kinematic blurring'' arguments due to many
line-of-sight structures are detailed in \citealp{kacprzak19}). Thus,
the kinematic signatures of outflows along the line-of-sight are more
likely to be preserved for lower mass galaxies than higher mass
galaxies.

Unlike the absorber kinematics, the absorber--galaxy kinematics
indirectly depend on halo mass. If mass was the most important galaxy
property governing the kinematics even after normalizing by $V_{\rm
  c}(D)$, then every panel in Figure~\ref{fig:mvazimuthorienovi} would
show significant differences between lower and higher mass
subsamples. However, the only significant differences found are for
comparisons in which outflows are expected to dominate the kinematics
of at least one of the subsamples. Overall, the lower mass face-on
galaxy subsample is the outlier, with velocity dispersions that are
significantly larger than both higher mass face-on galaxies
($6.7\sigma$) and lower mass edge-on galaxies ($3.3\sigma$). This is
consistent with the differential kinematics/wind recycling scenario
\citep{oppenheimer10, mathes14}.

\section{Summary and Conclusions}
\label{sec:conclusions}

We first examined the absorber kinematics for {\OVI} with the TPCF
method employed in \citet{magiicat5, magiicat4, nielsen17, magiicat6,
  pointon17}. The rest of the work analyzed absorber--galaxy
kinematics for {\OVI} gas using a modified version of the TPCFs. We
used a subset of 31 galaxies from the ``Multiphase Galaxy Halos''
Survey \citep{kacprzak15, kacprzak19, muzahid15, muzahid16, nielsen17,
  pointon17, pointon19} for the TPCF analysis; these are isolated
galaxies, where they have no neighboring galaxies within 200~kpc and
within a line-of-sight velocity separation of 500~{\kms}. The galaxies
span $0.12 < z_{\rm gal} < 0.66$, where $\sigma_z$ is better than
$\sim30~\kms$ \citep[e.g.,][]{kacprzak19} and are found within a
projected distance of 200~kpc from the quasar sightline. The galaxies
were imaged with ACS, WFC3 or WFPC2 on \emph{HST}, and the associated
morphological properties were modeled using GIM2D \citep{simard02,
  kacprzak15}.

We shifted the pixel velocities from the absorption spectra relative
to the galaxy systemic velocities, and normalized them with respect to
the circular velocity at the impact parameter, $V_{\rm c}(D)$, to
account for the range of halo masses ($10.9\leq\logtenmv\leq12.5$) in
the sample. We also found that this was crucial to eliminate any
potential mass biases in the results. We analyzed absorber--galaxy
kinematics of the subsamples derived from cuts by the halo mass,
$\logtenmv$, galaxy redshift, $z_{\rm gal}$, azimuthal angle, $\Phi$,
and inclination, $i$. The TPCFs are supplemented by average absorption
profiles, which provide optical depth distribution information. We also show the individual absorption profiles in the Appendix.

We thus find that:
\begin{enumerate}[nolistsep]
	\item There is a mass dependence in the {\OVI} absorber
          kinematics consistent with the column density--mass trends
          of \citet{oppenheimer16}. Lower mass galaxies
          ($\logtenmv<11.7$) tend to have narrower velocity
          dispersions than the typically more massive galaxies
          ($\logtenmv\sim12$). Subsamples with halo masses consistent
          with $L^{\ast}$ galaxies tend to have the largest velocity
          dispersions and this is reflected in both the TPCFs and the
          average absorption profiles. This mass subsample coincides
          with that of maximum {\OVI} column density due to having a
          virial temperature comparable to the temperature at which
          the {\OVI} ionization fraction is greatest. The largest halo
          masses, the group environment sample from \citet{pointon17},
          have the smallest velocity dispersions. This is likely due
          to more massive halos having a larger virial temperature,
          which provides the environment for {\OVI} to be promoted to
          higher order species (e.g., {\OVII} and beyond). The most
          massive halos are too hot for large {\OVI} ionization
          fractions, thus reducing the size of {\OVI} clouds and
          statistically leading to narrower kinematic extents.

        \item The total {\OVI} column densities for the lower mass,
          higher mass, and group samples are consistent with the
          aperture column densities from \citet{oppenheimer16}, for
          $\logtenmv\geq10.9$. This result provides further
          observational evidence of the virial temperature dependence
          of {\OVI}.
	
        \item After normalizing absorber--galaxy TPCFs by the circular
          velocity at the observed impact parameter, $V_{\rm c}(D)$,
          to account for halo mass, we found that lower mass galaxies
          ($\braket{\logtenmv}=11.64$) have significantly larger
          ($4.5\sigma$) absorber--galaxy velocity dispersions compared
          to higher mass galaxies ($\braket{\logtenmv}=12.06$). The
          average absorption profiles demonstrate that there is a
          larger fraction of gas with velocities greater than $V_{\rm
            c}(D)$ for the lower mass subsample, suggestive of
          outflowing gas \citep[e.g.,][]{oppenheimer10, mathes14,
            kacprzak19}.

        \item There are no significant differences for bivariate
          comparisons with subsamples sliced only by galaxy redshift
          ($2.6\sigma$), inclination ($0.01\sigma$), or azimuthal
          angle ($1.5\sigma$). However, these simple cuts neglect the
          fact that, e.g., gas along the projected major and minor
          axes of edge-on galaxies is expected to exhibit different
          kinematic signatures \citep[e.g.,][]{steidel02, ggk-sims,
            stewart11, bouche13}.

        \item A multivariate analysis investigating subsamples sliced
          by galaxy inclination angle and azimuthal angle shows
          potential outflow signatures. Large velocity dispersions and
          optical depths beyond $V_{\rm c}(D)$ are found for
          orientations in which outflows are expected to be most
          optimal. Face-on minor axis subsamples have the largest
          kinematic dispersions and a smoothly decreasing optical
          depth distribution out to large normalized velocities,
          reflecting outflowing gas that is likely decelerating and
          will eventually return to the galaxy. Edge-on minor axis gas
          has large optical depths that are concentrated near the
          galaxy systemic velocity, suggesting a large quantity of
          {\OVI}-absorbing gas being ejected perpendicular to the
          galaxy disk as expected for bipolar outflows. Accreting gas
          signatures are not directly observed for any subsample due
          to a combination of low covering fraction
          \citep[e.g.,][]{martin12, rubin-accretion} and ``kinematic
          blurring'' along the line-of-sight in which multiple
          kinematic structures are probed \citep[e.g.,][]{churchill15,
            peeples18,kacprzak19}.

        \item Combining the mass and galaxy orientation subsamples, we
          found that while outflows are most easily observed in lower
          mass galaxies, the clear outlying subsample is lower mass
          face-on galaxies. In this case, more recent star formation
          in the lower mass galaxies than would be expected in higher
          mass galaxies results in the {\OVI} gas having less time to
          mix kinematically to obscure the expected kinematic
          signatures of outflows. \\
 
\end{enumerate}

With these results, we suggest that in order to understand the physics
of {\OVI}-absorbing gas, it is imperative to first consider the halo
mass of the host galaxy. Combining the results of \citet{nielsen17}
and those presented in Section~\ref{sec:bivariate-abs}, {\OVI}
absorber kinematics largely represent the underlying virial
temperature of the host halo rather than baryon cycle processes,
unlike lower ions such as {\MgII} \citep{magiicat5,
  magiicat4}. Furthermore, accounting for the \citet{kacprzak19}
results for a subset of galaxies presented here, studying the circular
velocity-normalized absorber--galaxy kinematics of {\OVI} shows
indications of outflow signatures, but the kinematic blurring due to
multiple structures along the line-of-sight and the large cloud sizes
predicted for {\OVI} from photoionization models rule out easily
detecting accreting gas. Due to this, it is becoming increasingly
important for simulations to accurately model the CGM in order to
better understand the origins of the observed gas, and therefore
better understand how galaxies cycle their gas.

%%%%%%%%%%%%%%%%%%%%%%%%%%%%%%%%%%%%%%%%
\acknowledgments

We thank the anonymous referee, whose comments improved the clarity of the manuscript. M.N. acknowledges the support from the Vacation Scholarship program at
the Centre for Astrophysics \& Supercomputing at Swinburne University
of Technology. N.M.N. and G.G.K. acknowledge the support of the
Australian Research Council through a Discovery Project
DP170103470. S.K.P. acknowledges support through the Australian
Government Research Training Program Scholarship. Parts of this
research were supported by the Australian Research Council Centre of
Excellence for All Sky Astrophysics in 3 Dimensions (ASTRO 3D),
through project number CE170100012. S.M. acknowledges support from
European Research Council (ERC), Grant Agreement
278594-GasAroundGalaxies. C.W.C. and J.C.C. acknowledge the support
from award 1517831 from the National Science Foundation. J.C.C. is
also supported by National Science Foundation grant AST-1312686. This
work made use of Python packages NumPy and SciPy \citep{oliphant07},
Matplotlib \citep{hunter07}, IPython \citep{perez07}, tqdm
\citep{dacostaluis18}, Numba \citep{lam15}, and of NASA's Astrophysics
Data System Bibliographic Services.
 
Galaxy data presented here were obtained at the W. M. Keck
Observatory, which is operated as a scientific partnership among the
California Institute of Technology, the University of California, and
the National Aeronautics and Space Administration. The Observatory was
made possible by the generous financial support of the W. M. Keck
Foundation. Observations were supported by Swinburne Keck programs
2014A\_W178E, 2014B\_W018E, 2015\_W018E and 2016A\_W056E. The authors
wish to recognize and acknowledge the very significant cultural role
and reverence that the summit of Mauna Kea has always had within the
indigenous Hawaiian community. We are most fortunate to have the
opportunity to conduct observations from this mountain.

{\it Facilities:} \facility{Keck:II (ESI)}

%\newpage

\begin{appendix}

\section{Absorption Profiles for Absorber--Galaxy Kinematics}
\label{app:spectra}
  
\begin{figure*}[h]
	\includegraphics[width=\textwidth]{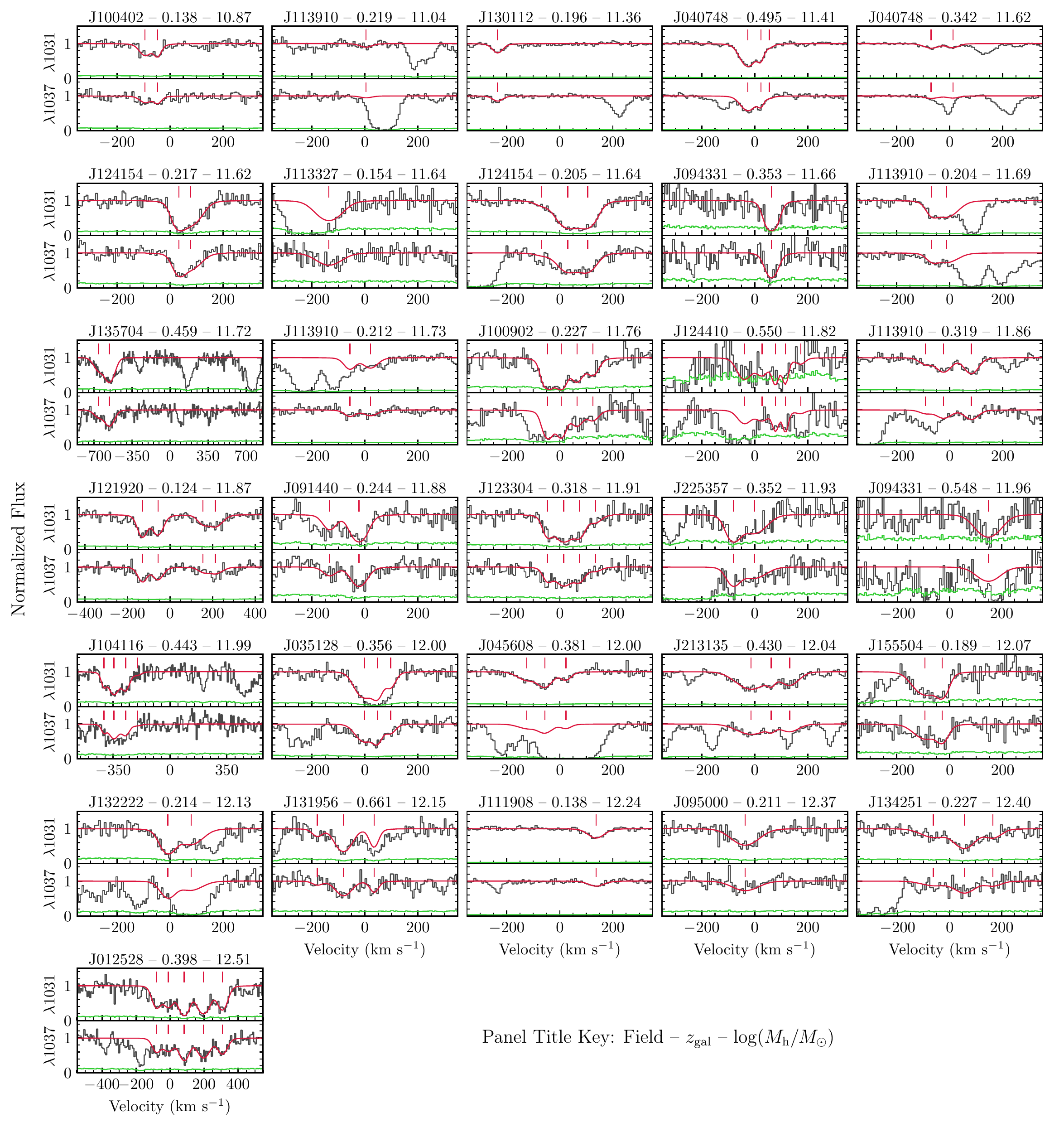}
	\caption[]{Absorption profiles (black histogram) of the {\OVI}
          $\lambda\lambda$1031,1037 doublet along with the
          corresponding uncertainties (green line), and fits (red
          line), for each sample absorber--galaxy pair. The pixel
          velocities in the absorption spectra have been shifted
          relative to the galaxy systemic velocity. The original
          absorption spectra can be found in \cite{nielsen17}. The
          {\OVI} $\lambda 1031$ and {OVI} $\lambda 1037$ lines are
          plotted on the top and bottom, in each pair of panels,
          respectively. The vertical red ticks at the top of each
          panel represent the central velocity of each Voigt profile
          component fitted to the data. Note that the limits on the
          velocity axes are not all identical.}
	\label{fig:spectra}
\end{figure*}

Figure~\ref{fig:spectra} presents the {\OVI} absorption profiles for
each absorber relative to the galaxy systemic redshift. These are the
profiles that are used for the absorber--galaxy kinematics analyses
presented in Section~\ref{sec:abs-galkin} \citep[for the absorbers
  used in the absorber kinematics analysis in
  Section~\ref{sec:bivariate-abs}, see][]{nielsen17}. Absorbers are
ordered by their host galaxy halo masses, with the field names,
redshifts, and halo masses labeled above each panel pair. These panel
pairs show both {\OVI} lines, with $\lambda 1031$ plotted on top and
$\lambda 1037$ on bottom. Data and uncertainties are plotted as black
and green histograms, respectively. Voigt profile models are plotted
as red curves while the individual fitted components are plotted as
red ticks.
 
\end{appendix}

%\newpage

\bibliography{references}
\bibliographystyle{apj}

\end{document}